\documentclass[reprint,amsmath,amssymb,onecolumn]{revtex4-1}
\usepackage{graphicx}	% Include figure files
\usepackage{dcolumn}	% Align table columns on decimal point
\usepackage{bm}		% bold math
\usepackage{color}
\usepackage{docmute}
\usepackage{float}
\usepackage{xr}
\begin{document}
\renewcommand \thetable{\arabic{table}}
\title{The study of magnetic topological semimetals by first principles calculations}
\author{Jinyu Zou}
\author{Zhuoran He}
\author{Gang Xu}
 \email{gangxu@hust.edu.cn}
\affiliation{Wuhan National High Magnetic Field Center $\&$ School of Physics, Huazhong University of Science and Technology, Wuhan 430074, China}

\begin{abstract}
Magnetic topological semimetals (TSMs) are topological quantum materials with broken time-reversal symmetry (TRS) and isolated nodal points or lines near the Fermi level. Their topological properties would typically reveal from the bulk-edge correspondence principle as nontrivial surface states such as Fermi arcs or drumhead states, etc. Depending on the degeneracies and distribution of the nodes in the crystal momentum space, TSMs are usually classified into Weyl semimetals (WSMs), Dirac semimetals (DSMs), nodal-line semimetals (NLSMs), triple-point semimetals (TPSMs), etc. In this review article, we present the recent advances of magnetic TSMs from a computational perspective. We first review the early predicted magnetic WSMs such as pyrochlore iridates and HgCr$_2$Se$_4$, as well as the recently proposed Heusler, Kagome layers, and honeycomb lattice WSMs. Then we discuss the recent developments of magnetic DSMs, especially CuMnAs in Type-III and EuCd$_2$As$_2$ in Type-IV magnetic space groups (MSGs). Then we introduce some magnetic NLSMs that are robust against spin-orbit coupling (SOC), namely Fe$_3$GeTe$_2$ and LaCl (LaBr). Finally, we discuss the prospects of magnetic TSMs and the interesting directions for future research.
\end{abstract}

\maketitle

\section*{\label{intro} Introduction}

The classification of material phases and description of phase transitions in condensed matter physics have long been given by the Landau theory of spontaneous symmetry breaking, with different phases described by different local order parameters. People could understand, for example, the superconducting phase transition from the breaking of the $U(1)$ gauge symmetry, the ferromagnetic phase transition from the breaking of the time-reversal symmetry (TRS) and all sorts of structural phase transitions in crystals from the change of space group symmetries. Despite the great triumph of Landau theory, its limitations reveal when Klitzing discovered the quantum Hall effect (QHE) in a 2D electron gas (2DEG) under high magnetic fields \cite{Klitzing1980}. This remarkable discovery then opened a new field of study for the phase transitions of materials, i.e., the so-called topological phase transitions. QHE is beyond Landau theory because the transitions between electronic states holding different integer Hall conductances do not break any symmetry. Thouless \textit{\textit{et al.}}$\;\!$\cite{Thouless1982} used the Kubo formula to calculate and interpret the integers and found their topological origin. These topological integers are known as the TKNN numbers in memory of their pioneering works and are now understood as the first Chern number in topological band theory.

The QHE in a 2DEG was later reproduced in the topological phases of a 2D lattice by Haldane, who proposed a honeycomb lattice model without applying any net magnetic fields \cite{Haldane1988}. The TRS was broken by the staggered magnetic fluxes over the lattice with zero total flux. This was the first model for the quantum anomalous Hall effect (QAHE), where the quantized Hall conductance was characterized by the first Chern number but realized with no magnetic field applied. The first Chern number is calculated by the integral of the Berry curvature of the occupied bands over the first Brillouin zone (BZ) divided by $2\pi$, and is stable against smooth perturbations of the system without closing the band gap. Such a topological invariant can only be defined for even-dimensional systems and is only nonzero for magnetic systems where the TRS is broken.

Nearly 20 years later, Kane and Mele proposed a new topological invariant \cite{Kane2005} --- the $Z_2$ number. They also studied a 2D honeycomb lattice model but with TRS preserved and spin-orbit coupling (SOC) considered. The model has vanishing QHE, although the quantum spin Hall effect (QSHE) can be realized. The $Z_2$ topological number of the model is characterized by the difference of the Chern numbers of spin-up and spin-down states modulo 2, which can be either 0 (trivial) or 1 (nontrivial). Such a topological classification was then generalized to 3D systems to describe the nontrivial band insulators \cite{Fu2007,Fu2011,Hasan2010,Qi2011}, which are known as the 3D topological insulators (TIs). These pioneering models and theoretical works inspired the prosperity of theoretical predictions and experimental realizations of the topologically nontrivial materials later on. The QSHE system of HgTe/CdTe quantum well was predicted \cite{Bernevig2006} and soon confirmed experimentally \cite{Konig2007}. 3D TIs were discovered in the Bi$_2$Se$_3$ family \cite{ZhangHaijun2009,Xia2009,Chen2009}, and QAHE was predicted and observed in the magnetically doped thin films of the Bi$_2$Te$_3$ family \cite{Yu2010,Chang2013}.

The topological classification can be generalized to semimetals, known as topological semimetals (TSMs), where the lowest conduction band and highest valence band cross each other at isolated points (nodes) or lines (nodal lines) at the Fermi level \cite{Fang2016,Yan2017,Burkov2018,Armitatge2018}. In the beginning, TSMs were mainly discussed as an intermediate phase between normal insulators (NIs) and TIs. When the inversion symmetry (IS) is broken, gapless points can appear in pairs during the NI-to-TI transition and move in the Brillouin zone under continuous changes of the model parameters until they meet and annihilate as the system reaches a TI phase \cite{Murakami2007,Murakami2007a,Murakami2008}. The intermediate gapless phase is called a Weyl semimetal (WSM) because the low-energy excitations near a two-fold degenerate point, called a Weyl point (node), are linearly dispersive and satisfy the Weyl equation that describes massless Weyl fermions in high-energy physics. Also, The WSM phase was modeling studied by alternately stacking thin films of magnetically doped TIs and NIs \cite{Burkov2011,Balents2011}. At the same time, the single-crystal WSM candidates such as pyrochlore iridates \cite{Wan2011} and HgCr$_2$Se$_4$ \cite{Xu2011HgCr2Se4} were predicted. These early works of WSMs stimulated the research interest for TSMs greatly.

According to the Nielsen-Ninomiya theorem \cite{Nielsen1981,Nielsen1981a}, Weyl nodes always appear in pairs. They are topologically stable because the Weyl Hamiltonian near a Weyl node
\begin{align}
H(\vec{k})=\sum_{ij}A_{ij}k_i\sigma_j,\quad i,j=1,2,3,
\end{align}
has used up all the three Pauli matrices $\sigma_1,\sigma_2,\sigma_3$. Perturbations can only move the Weyl node in the crystal momentum space but cannot annihilate it unless it meets with another Weyl node holding opposite chirality and opens a band gap. The chirality here is defined as the Chern number of the Bloch states on a 2D spherical surface enclosing the Weyl node. The result is given by
\begin{align}
C=\mathrm{sgn}\,_{\!}[\det(A)]=\pm 1,
\end{align}
assuming the $3\times 3$ matrix $A$ has full rank so there is no nodal-line direction in BZ. Hence, a Weyl node is like a magnetic monopole in the crystal momentum space and can behave like either a ``source'' ($C=+1$) or a ``sink'' ($C=-1$) of the Berry curvature.

Analogous to WSMs, we also have Dirac semimetals (DSMs) with a four-fold degenerate Dirac node and the low-energy excitations near the node satisfy the four-component massless Dirac equation
\begin{align}
i\hbar\gamma^\mu\partial_\mu\psi=0,\quad \mu=0,1,2,3,
\end{align}
with $4\times 4$ matrices $\gamma^0 = \tau_3 \otimes I_{2\times 2}$ and $\gamma^j = i\tau_2 \otimes \sigma_j$, $j=1,2,3$ in the standard representation. Here we use two sets of Pauli matrices $\mathbf{\tau}$ and $\mathbf{\sigma}$ to distinguish the direct-product spaces. The Dirac Hamiltonian can be rewritten as $H(\vec{k})=\hbar c\,_{\!}\tau_1\otimes \vec{k}\cdot\vec{\sigma}$, with the Pauli vector $\vec{\sigma}=(\sigma_1,\sigma_2,\sigma_3)$. Since $H(\vec{k})$ commutes with the $\gamma^5$-symmetry operator
\begin{align}
\gamma^5 = i\gamma^0 \gamma^1 \gamma^2 \gamma^3 = \tau_1 \otimes I_{2\times 2},
\end{align}
with eigenvalues $\gamma^5=\pm 1$, the 4-dimensional Hilbert space of $\psi$ can be reduced into two uncoupled two-dimensional subspaces of Weyl fermions with effective Hamiltonians $H(\vec{k})=\pm \hbar c\,_{\!}\vec{k}\cdot\vec{\sigma}$, respectively. Hence, a Dirac node can be viewed as a four-fold degenerate ``kissing'' point of two Weyl nodes with opposite chiralities. It is not topologically stable against the mass term, which breaks the $\gamma^5$-symmetry and couples the two Weyl subspaces to open a gap. In order to obtain a stable Dirac node, additional crystalline symmetries are necessary to protect the Dirac nodes on the high-symmetry points or lines in the first BZ \cite{Yang2014}. Such kind of DSM states have been theoretically predicted \cite{Wang2012,Wang2013} and experimentally confirmed in nonmagnetic materials Na$_3$Bi \cite{Liu2014864} and Cd$_3$As$_2$ \cite{Borisenko2014,Neupane2014}, etc.

Beyond the homologous particles such as the Weyl fermions and Dirac fermions, there are also other types of quasiparticles that are allowed in solids by the representation theory of crystalline symmetries, which are the so-called new fermions \cite{Bradlyn2016Science}. As pointed out by Bradlyn \textit{et al.}, the irreducible representations of the little group at the high-symmetry BZ points in some specific space groups could suggest three-fold \cite{Fang2012,Zhu2016,Chang2017a}, six-fold \cite{Chang2017} or eight-fold \cite{Wieder2016,Schoop2018,Geilhufe2017} band degeneracy. A three-fold degenerate node can also be formed by one two-fold degenerate band and one single band on the high-symmetry lines of BZ \cite{Weng2016,Cheung2018}. The semimetals holding three-fold degenerate nodes on the high-symmetry points or lines at the Fermi level are both called triple-point semimetals (TPSMs). The semimetals with eight-fold degenerate nodes are also called double Dirac semimetals (DDSMs), just like two overlapping Dirac nodes in the crystal momentum space.

If the valence and conduction bands are not touching at isolated points in the crystal momentum space, but at continuous one-dimensional Fermi lines (including loops, chains and links) at the Fermi level, the semimetal is called a nodal-line semimetal (NLSM) \cite{Fang2016,Burkov2011PRB,Fang2015,Bian2016,Hu2016}. A nodal line can be viewed as a ``kissing'' line of Weyl nodes or Dirac nodes with a Chern monopole charge (for Weyl nodal lines) or a $Z_2$ monopole charge (for Dirac nodal lines). Nodal lines are generally unstable against perturbations but can be protected by crystalline symmetries at high-symmetry planes (e.g., mirror planes) in the Brillouin zone.

Ever since the first theoretical prediction of pyrochlore iridates as candidates of WSMs, topological semimetals have become a highly attractive field of study. Currently, most of the TSMs calculated from {first principles} calculations and studied experimentally are nonmagnetic, i.e., the TRS-preserved semimetals, including the well known TaAs family \cite{TaAsFamily,Lv2015,Shekhar2015}, Cd$_3$As$_2$ \cite{Wang2013,Liu2014}, Na$_3$Bi \cite{Wang2012,Liu2014864}, etc. In recent years, magnetic TSMs are receiving more and more attention, as they have several advantages over nonmagnetic TSMs. First, some magnetic WSMs can host only one pair of Weyl nodes, which are ideal for transport and chiral anomaly studies. Second, systems with broken TRS can have nonzero net Berry curvatures, which can induce unique properties such as intrinsic anomalous Hall effect, thermoelectric currents (anomalous Nernst effect), etc. Third, the half-metallic feature of some magnetic TSMs such as HgCr$_2$Se$_4$, Heusler compounds and Co$_3$Se$_2$S$_2$ makes them good for spin manipulations and spintronics applications. Finally, the magnetic materials are more varied and richer, and the magnetic space group  (MSGs) are much larger and complex than space groups, which may derive some novel magnetic TSMs.

According to group theory \cite{Bradley2010b}, there are 1651 magnetic space groups (MSGs), which are divided into four types:
\begin{subequations}
\begin{align}
M_1&=G,\\
M_2&=G\cup\mathcal{T}G,\\
M_3&=G\cup R\mathcal{T}G,\\
M_4&=G\cup\tau\mathcal{T}G.
\end{align}
\end{subequations}
Here $G$ is the unitary subgroup of $M_i,\,i=1,2,3,4$, which is an ordinary crystalline space group, $\mathcal{T}$ is the antiunitary time-reversal operator, $R\notin G$ is a Euclidean symmetry other than a pure translation, and $\tau\notin G$ is a translation operator connecting the spin-up and spin-down sublattices. There are 230 ordinary crystalline space groups of Type I, 230 TRS-preserved space groups (i.e.~the gray MSGs) of Type II, 674 MSGs of Type III and 517 MSGs of Type IV.

As we discussed above, the first theoretically predicted TSMs were magnetic pyrochlore iridates and HgCr$_2$Se$_4$. Recently, more and more magnetic WSMs have been proposed, including Heusler compounds \cite{Wang2016,Kubler2016,Chang2016}, {Kagome} layers \cite{Yang2017,Liu2018,Wang2018} and honeycomb-lattice materials \cite{Nie2017}. The study of magnetic DSMs has also made great progress in the past several years. Candidates of DSMs in Type-III and Type-IV MSGs were proposed, namely CuMnAs \cite{Tang2016} and EdCd$_2$As$_2$ \cite{Hua2018}. Recently, SOC-robust magnetic NLSMs were predicted to emerge in the layered system Fe$_3$GeTe$_2$ \cite{Kim2018} and LaCl (LaBr) \cite{Nie2019} from {first principles} calculations. Compared with these theoretical advances, the experimental studies of magnetic TSMs have been rarer and harder. The main difficulty comes from three aspects: a) many magnetic TSMs proposed are metastable, which makes their high-quality crystal samples difficult to synthesize, and b) their topological properties can highly depend on their magnetic configuration and magnetic moment direction, which may get mispredicted sometimes by {first principles} calculations, and c) their complicated domain walls often make their topological band structures difficult to measure and confirm using current experimental techniques such as angle-resolved photoemission spectroscopy (ARPES).

In this review article, we will mainly focus on the recently proposed magnetic TSMs from {first principles} calculations. Section ``Magnetic Weyl Semimetals'' presents the proposed candidates of magnetic WSMs in chronological order. Section ``Magnetic Dirac Semimetals'' reviews the predicted magnetic DSMs CuMnAs and EuCd$_2$As$_2$ with Type-III and Type-IV MSG symmetries, respectively. Section ``Magnetic Nodal Line Semimetals'' reviews the magnetic NLSMs Fe$_3$GeTe$_2$ and LaCl (LaBr), which are predicted to be robust against SOC on certain conditions. In the last section, we discuss the potential applications and possible future research directions of magnetic TSMs.

\section*{\label{MWSM} Magnetic Weyl Semimetals}
Weyl semimetals (WSMs) are generally divided into two types: magnetic WSMs and noncentrosymmetric WSMs, which correspond to the breaking of the time-reversal symmetry $\mathcal{T}$ and the inversion symmetry $I$, respectively. If both symmetries $\mathcal{T}$ and $I$ are preserved, the two Weyl nodes with opposite chiralities will meet at the same $\vec{k}$-point to form a Dirac node. So to create Weyl nodes, either $\mathcal{T}$ or $I$ needs to be broken. Historically, the first types of theoretically predicted topological semimetals were magnetic WSMs in pyrochlore iridates with strong spin-orbit coupling and all-in/all-out (AIAO) magnetic configurations \cite{Wan2011}.
{Magnetic WSMs are not sufficiently studied at present because of the experimental difficulties due to their complex domain structures.} However, magnetic WSMs are worth studying due to their unique properties such as large intrinsic anomalous Hall effect (AHE) and anomalous Nernst effect (ANE), which can be useful for building electronic devices.
The AHE is related to the integral of the Berry curvature of the occupied bands in the BZ \cite{Fang2003,Haldane2004,Xiao2006,Wang2007,Nagaosa2010,Xiao2010,Gradhand2012}, and is only possible in magnetic materials. More explicitly, the intrinsic zero-temperature Hall conductivity at Fermi energy $E_F$ is expressed as \cite{Xiao2006}
\begin{equation}\label{AHE}
  \sigma_{ij}(E_F) = -\frac{e^2}{\hbar}\int [dk]\,_{\!}\Theta(E_F-\epsilon_k)_{\,}\Omega_l(k)
\end{equation}
where $i,j,l=x,y,z$, $\Theta$ is the step function and $\Omega_l$ is $l$ component of the Berry curvature. The Berry curvature is highly enhanced near Weyl nodes, making large AHE possible in magnetic WSMs if the Fermi level is close to the Weyl nodes \cite{Burkov2011,Yang2011,ChenPRB2013,Burkov2014}. Moreover, the carrier density is reduce to zero at the Weyl nodes, which suggests large anomalous Hall angle in those materials. The ANE is a nontrivial thermoelectric phenomena where a temperature gradient and a perpendicular magnetization can induce a transverse electric voltage \cite{Lee2004,Miyasato2007,Yong2008}. As the Berry curvature behaves like a magnetic field, like the AHE, the thermoelectric conductivity can also be calculated by an integral of the Berry curvature \cite{Xiao2006,Zhang2009,Gradhand2012,Dumitrescu2012}, which then gives rise to the Mott relation
\begin{equation}\label{ANE}
\alpha_{ij} = \frac{\pi^2}{3}\frac{k_B^2T}{e}\,_{\!}\sigma_{ij}'(E_F),
\end{equation}
where $\sigma_{ij}'(E_F)$ is the energy derivative of the intrinsic anomalous Hall conductivity. Thus one immediately expects that a giant ANE can also be generated in magnetic WSMs \cite{Sharma2016,Ikhlas2017,Li2017,Sakai2018,Noky2018}.
We will review some typical magnetic WSMs in this section, such as pyrochlore iridates, HgCr$_2$Se$_4$, Heusler compounds, Kagome layers and honeycomb lattice GdSI.

\subsection*{Pyrochlore iridates}

In 2011, Wan et al \cite{Wan2011} first reported that the $5d$ transition metal oxides pyrochlore iridates A$_2$Ir$_2$O$_7$ (A=Y or rare-earth element) with AIAO magnetic order can be turned into the WSM phase. By the method of a ``plus $U$'' extension of density functional theory (DFT+$U$), they found 24 Weyl nodes in bulk and abundant Fermi arcs on surface at intermediate electronic correlation $U \sim 1.5$~eV.

\begin{figure}[!h]
  \centering
  \includegraphics[width=0.7\linewidth]{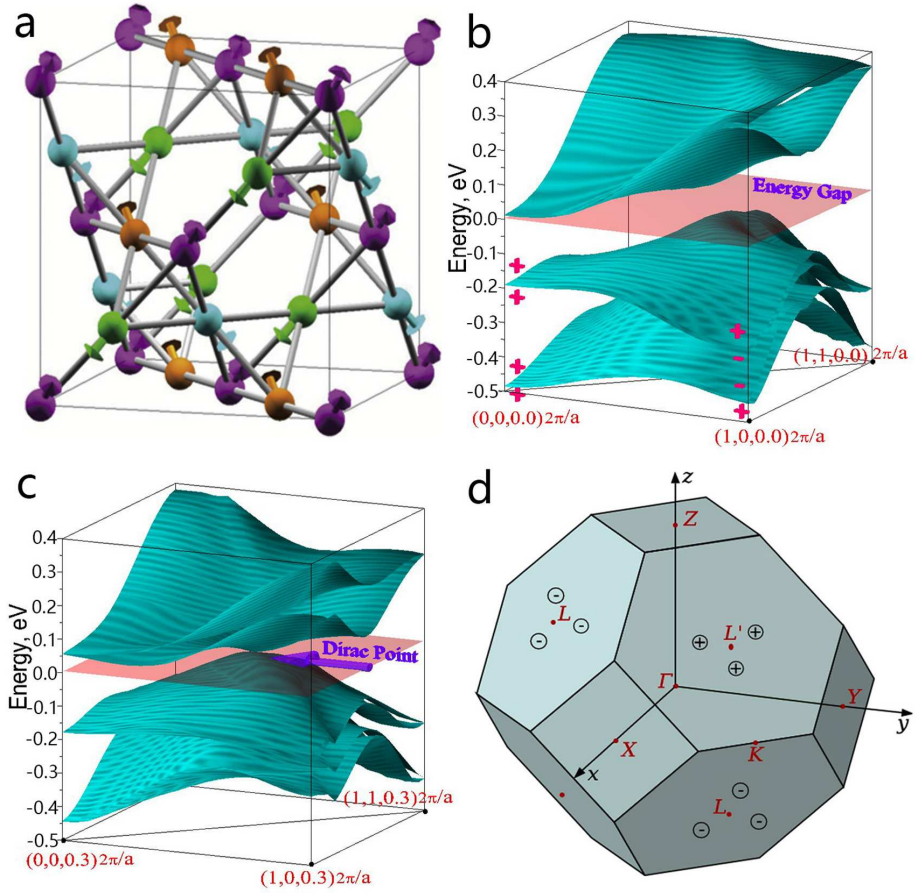}\\
  \caption{(a) The crystal structure of pyrochlore iridates. Ir atoms are located at the corner of the tetrahedral network, and their magnetic moments are predicted to form a AIAO configuration. (b-d) The band structure and Weyl nodes in WSM phase at $U=1.5$~eV calculated by the LSDA+$U$+SO method. (b) The energy bands with in the $K_z = 0$ plane, where ($\pm$) denotes the band parities; (c) The energy bands and a Weyl node in the $k_z = 0.6\pi/a$ plane; (d) The Weyl nodes in the three-dimensional Brillouin zone. Nine of the 24 nodes are shown here, the others are related by three fold rotation symmetry and IS. The circled ($\pm$) denotes the chirality.\\
  (\textbf{a} is reprinted from ref. \cite{WanReview2016}, Computational Materials Science. Copyright \copyright2019 by Elsevier.\\
  \textbf{b-d} are reprinted from ref. \cite{Wan2011}. Copyright \copyright2019 by the American Physical Society.)} \label{Fig1}
\end{figure}

The calculations show that the influence from the rare-earth element on the bands near Fermi level is negligible in A$_2$Ir$_2$O$_7$, therefore, Wan et al focus on Y$_2$Ir$_2$O$_7$ to discuss the magnetic configuration and topological phase transition. In the pyrochlore iridates crystal, the corner-sharing tetrahedra of Ir sublattice is largely geometrically frustrated, and the calculation gives the AIAO magnetic configuration, see Fig.~\ref{Fig1}a. Ir$^{4+}$ is located at the tetrahedra corner with $5d^5$ outer-shell electrons half-filling the ten $d$ levels. The surrounding oxygen octahedra provides a large crystal-field and causes the splitting between the doubly degenerated $e_g$ and triply degenerated $t_{2g}$ states. $e_g$ bands are about $2$~eV higher, hence the Fermi level is mainly dominated by $t_{2g}$ bands. Due to the strong SOC of $5d$ transition metal element, the $t_{2g}$ states further split to higher $J=1/2$ doublet and lower $J=3/2$ quadruplet. The latter is fully filled as Ir$^{4+}$ has five $d$-{electrons}, $J=1/2$ doublet is half-filled and mainly dominate the low energy properties of band structure. Hence there are eight half-filling bands near Fermi level given the four Ir atoms in each unit cell. On the other hand, the electron correlation effect can not be ignored. considering the correlation $U$, local spin density approximation (LSDA)+ SO + $U$ calculations show the phase transition from normal metal at small $U$ to WSM at intermediate $U \sim 1.5$~eV and Mott insulator phase at $U$ above $2$~eV.

In the weak correlation limit, the band structure of non-magnetic phase calculated by LDA + SO method (without $U$) reveals that the eight levels near Fermi energy are in the sequence 2,4,2 of degeneracies, which must be metallic phase in the half-filling. On the contrary, experiments show that Y$_2$Ir$_2$O$_7$ is an insulator \cite{Taira2001,Fukazawa2002,Soda2003}. Considering $U$ and other magnetic configuration still can not open the gap, but an insulation band structure can be obtained in strong correlation limit ($U > 1.8$~eV) and AIAO order, known as Mott insulator. At the intermediate correlation $U\sim 1.5$~eV, as shown in Fig.~\ref{Fig1}b-c, the band structure of AIAO magnetic order calculated by LSDA + SO + $U$ demonstrates 24 Weyl nodes in BZ related by three fold rotation symmetry (same chirality) and IS (opposite chirality). Because of symmetry, all Weyl nodes are in the same energy. Adjusting $U$ can move the Weyl nodes. With $U$ increasing, Weyl nodes can move to meet at L point and annihilate, driving to a Mott insulator phase. With $U$ decreasing to around $1$~eV, two opposite Weyl nodes can annihilate at X point, and Wan et al suggested axion insulator phase may appear. Unfortunately, the material will transform to a metallic phase around $U\sim 1$~eV before the the Weyl nodes annihilate according to the band structure calculation.

The AIAO ground-state magnetic configuration, which is originated from the nearest-neighbor antiferromagnetic coupling and strong geometric frustration of the pyrochlore lattice, has been experimentally confirmed \cite{Disseler2012,Tomiyasu2012,Disseler2014,Lefran2015}. The magnetic frustration, electronic correlation and strong SOC of the 5d orbitals in transition metal elements are crucial for understanding the origin of the WSM phase in pyrochlore iridates, and are also a treasury of other topological phenomena such as topological insulators, axion insulators and topological Mott insulators \cite{Witczak-Krempa2014,WanReview2016}. Witczak-Krempa \textit{et al.}~\cite{Witczak-Krempa2012,Go2012,Witczak-Krempa2014} established a minimal model with the Hubbard Hamiltonian to capture the magnetic ground states and the topological phase by changing correlation $U$.
Although the theoretical prediction of this magnetic WSM phase have not been directly confirmed by experiment, the study on pyrochlore iridates through varies of theoretical methods \cite{Chen2012,Moon2013,Wang2017,Wang2017b,Berke2018} and indirect experimental signals \cite{Sushkov2015,Ueda2016,Nakayama2016} is lasting to shed light on the Weyl nodes and their stabilities.
For example, the discovery of the conducting magnetic domain walls in the insulating bulk pyrochlore iridates \cite{Moccia2014,Ma2015} can be explained as the surviving mid-gap states at the domain wall \cite{Yamaji2014}.

\subsection*{HgCr$_2$Se$_4$}

\begin{figure}[!h]
  \centering
  \includegraphics[width=0.7\linewidth]{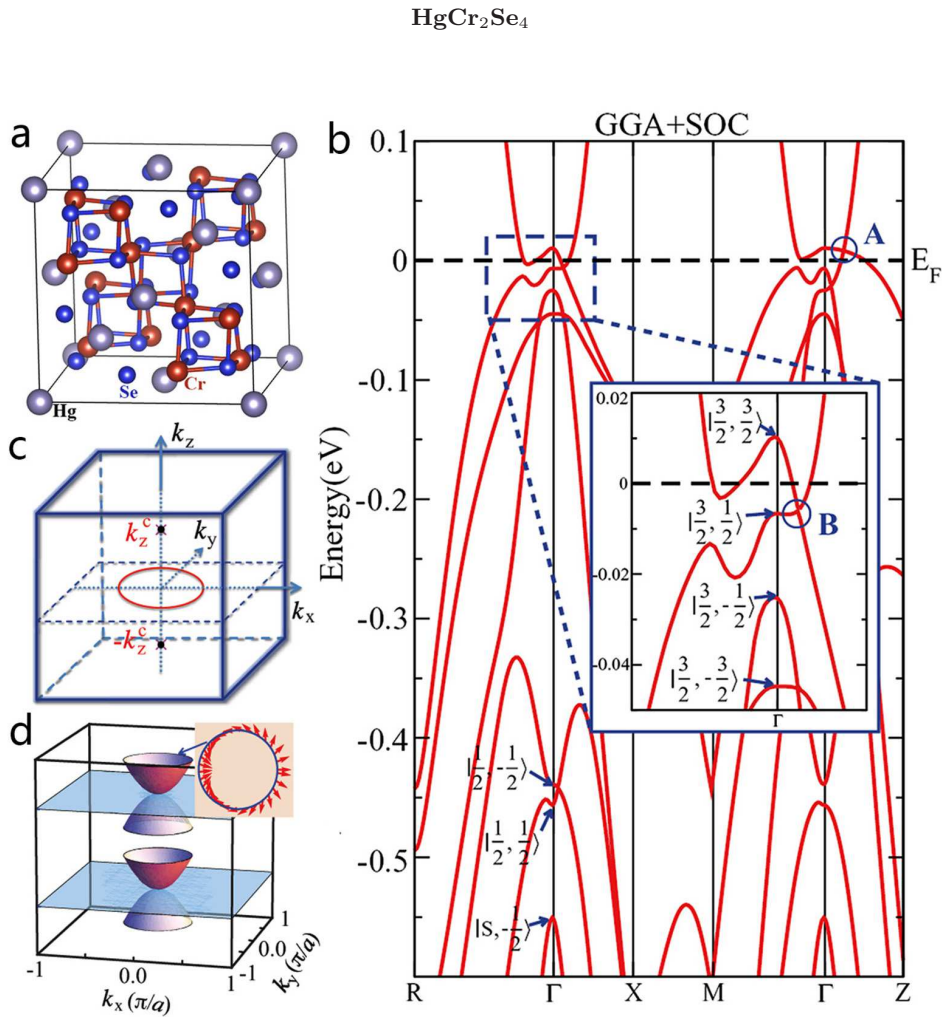}\\
  \caption{(a) Crystal structure of HgCr$_2$Se$_4$ spinel. (b) The band structure with SOC, where the majority spin aligns to the (001) direction. (c-d) Weyl nodes and gauge flux in HgCr$_2$Se$_4$. (c) Two Weyl nodes located on the $k_z$ axis; (d) The schematic plot of the band dispersion around the Weyl nodes in the $k_z=\pm k_z^c$ plane, the inset shows the chiral spin texture.\\
  (\textbf{a} is reprinted from ref. \cite{Weng2016review}. \copyright2016 IOP Publishing Ltd; permission conveyed through Copyright Clearance Center, Inc.\\
  \textbf{b-d} are reprinted from ref. \cite{Xu2011HgCr2Se4}. Copyright \copyright2019 by the American Physical Society.) }\label{Fig2}
\end{figure}
The Weyl nodes in pyrochlore iridates are subtle and sensitive to the fine-tuning of the electronic correlation $U$. Also there are many Weyl nodes in BZ, making it complicated to analyze the WSM phase. Nearly at the same time, Xu \textit{et al.}~\cite{Xu2011HgCr2Se4} proposed the ferromagnetic material HgCr$_2$Se$_4$ with only one single pair of Weyl nodes with chirality $\pm 2$. HgCr$_2$Se$_4$ is a ferromagnetic spinel exhibiting large coupling effects between electronic and magnetic properties \cite{Wojtowicz1969}.
The spinel structure, with space group $Fd\bar{3}m$, can be related to the diamond structures by taking the small Cr$_2$Se$_4$ cluster as a single pseudo-atom (called X) located at the center of mass, see Fig.~\ref{Fig2}(a), therefore Hg and X form two embedded diamond structure. The Cr$_2$Se$_4$ cluster are connected by the corner sharing Cr atoms, hence each Cr atom is octahedrally coordinated by the 6 nearest Se atoms.

The {first principles} calculation confirms the ferromagnetic order with a total energy about $2.8$~eV/f.u.~lower than the nonmagnetic phase. The obtained magnetic moment ($6.0\mu_B/f.u.$) agrees with experiments \cite{Baltzer1965,Baltzer1966} very well. Without SOC, it is suggested that the system can be approximately characterized as a ``zero-gap half-metal''. It is a half-metal because of the presence of a gap in the spin-up channel and it is zero-gap because of the band-touching around the $\Gamma$ point in the spin-down channel. The Cr$^{3+}$ $3d$ states are strongly spin-polarized, resulting in the configuration $t_{2g}^{3\uparrow}e_g^{0\uparrow}t_{2g}^{0\downarrow}e_g^{0\downarrow}$. The octahedral crystal field surrounding the Cr atoms is strong and opens a gap between the $t_{2g}^{3\uparrow}$ and $e_g^{0\uparrow}$ subspaces. The top of the valence band from $-6$ to $0$~eV is dominated by Se-$4p$ states. Due to the hybridization with Cr-$3d$ states, Se-$4p$ are slightly spin-polarized but with an opposite moment (about $-0.08\mu_B$/Se). The zero-gap behavior in the down spin channel is the most important character, which suggests a band inversion around $\Gamma$, similar to the case in HgSe or HgTe \cite{Delin2002,Moon2006}.

The four low energy states (8 after considering spin) at the $\Gamma$ point are the linear combinations  $|P_x\rangle,|P_y\rangle,|P_z\rangle,|S\rangle$, with $|P_\alpha\rangle \approx \frac{1}{\sqrt{8}}\sum_{i=1}^8 |p_{\alpha}^i\rangle$ and $|S\rangle \approx 0.4\sum_{j=1}^2|s^j\rangle + 0.24\sum_{k=1}^4|d_{t_{2g}}^k\rangle$, where $\alpha = x, y, z$ and i, j, k respect Se, Hg, Cr atoms, $|s\rangle,|p_{\alpha=x,y,z}\rangle,|d_{t_{2g}=xy,yz,zx}\rangle$ are corresponding atomic orbits of each atom. Taking these four states as bases, one can found the same situation as in HgSe and HgTe, the only difference is the presence of exchange splitting. The band inversion, where $|S, \downarrow\rangle$ being lower than $|P,\uparrow\rangle$ at $\Gamma$ point, is due to the following two factors. Firstly, the Hg-5d states are very shallow [located at about $-7.0$~eV] and its hybridization with Se-4p states will push the anti-bonding Se-4p states higher, similar to HgSe. Secondly, the hybridization between unoccupied Cr-3d$^\downarrow$ and Hg-6s$^\downarrow$ states will push the Hg-6s$^\downarrow$ state lower in energy. Thus the $|S, \downarrow\rangle$ is about $0.4$~eV lower than the $|P, \downarrow\rangle$ states, and further enhanced to be $0.55$~eV in the presence of SOC. One should be aware of the correlation effect beyond GGA, because the higher the Cr-3d$^\downarrow$ states, the weaker the hybridization with Hg-6s$^\downarrow$. It has been proved that the LDA + $U$ calculations with effective $U$ around $3.0$~eV can describe the semiconducting CdCr$_2$S$_4$ and CdCr$_2$Se$_4$ very well \cite{Fennie2005,Yaresko2008}. As for HgCr$_2$Se$_4$, The same LDA + $U$ calculations shows that the band inversion remains unless the $U$ is unreasonably large ($> 8.0$~eV).

When considering SOC, the new low-energy states at $\Gamma$ are $|\frac{3}{2},\pm\frac{3}{2}\rangle, |\frac{3}{2},\pm\frac{1}{2}\rangle, |\frac{1}{2},\pm\frac{1}{2}\rangle$, and $|S,\pm \frac{1}{2}\rangle$ contribute from $|P\rangle$ and $|S\rangle$ states. The exchange splitting energetically separates the eight bands, with the highest $|\frac{3}{2},\frac{3}{2}\rangle$ and lowest $|S,-\frac{1}{2}\rangle$ state. Several band crossings can be observed in the band inversion, as shown in Fig.~\ref{Fig2}(b). Among them, however, only two kinds of band crossings (called A and B) are important for the states very close to the Fermi level. The crossing A gives two points located at $k_z=\pm k_z^c$ along the $\Gamma - Z$ line, the trajectory of crossing B is a closed loop surrounding the $\Gamma$ point in the $k_z=0$ plane, as schematically shown in Fig.~\ref{Fig2}(c). Given a 2D plane with fixed $k_z $($k_z\neq 0$ and $k_z \neq \pm k_z^c$), the band structure are all gapped, hence one can calculate its Chern number $C$. It turns out that $C=0$ for the planes with $k_z<-k_z^c$ or $k_z>k_z^c$, while $C=2$ for the planes with $-k_z^c<k_z<k_z^c$ and $k_z\neq 0$. Hence the crossing A points locate at the phase boundary between $C = 2$ and $C = 0$ planes are topologically unavoidable Weyl nodes. On the other hand, the crossing B points, i.e. the closed loop in the $k_z = 0$ plane is a Weyl nodal line due to the mirror symmetry. Therefore, HgCr$_2$Se$_4$ is a material with coexisting Weyl points and Weyl nodal lines when the crystal mirror symmetry is preserved.

To capture the band inversion nature of $|\frac{3}{2},\frac{3}{2}\rangle$ and $|S,-\frac{1}{2}\rangle$ at $\Gamma$ point, one can downfold the $8\times8$ $\mathbf{k}\cdot\mathbf{p}$ effective Hamiltonian to a $2\times 2$ model:
\begin{equation}\label{HgCr2Se4_eq1}
  H_\mathrm{eff} = \left(
              \begin{array}{cc}
                M & Dk_zk_-^2 \\
                Dk_zk_+^2 & -M \\
              \end{array}
            \right),
\end{equation}
where $k_{\pm} = k_x \pm ik_y$, $M = M_0 - \beta k^2$ is the mass term expanded to the second order, and $M_0 > 0, \beta > 0$ to ensure the band inversion. The two bases have opposite parity, hence the off-diagonal element has to be odd in k. $k_{\pm}^2$ is to conserve the angular momentum along z direction. Thus, to the leading order, $k_zk_{\pm}^2$ is the only possible form for the off-diagonal element. The energy eigenvalues $E(k)=\pm \sqrt{M^2 + D^2k_z^2(k_x^2+k_y^2)^2}$ suggest two gapless solutions: one is the degenerate points along $\Gamma - Z$ line with $k_z = \pm k_z^c = \pm\sqrt{M_0/\beta}$ ; the other is a circle around $\Gamma$ point in the $k_z = 0$ plane determined by the equation $k_x^2 + k_y^2 = M_0/\beta$. They are exactly consistent with the {first principles} calculation. The dispersion of two Weyl nodes are quadratic rather than linear, with their chirality are $\pm 2$ respectively, and the Chern number $C=2$ for the planes with $-k_z^c < k_z < k_z^c$ and $k_z \neq 0$. Two opposite Weyl nodes form a single pair of magnetic monopoles carrying the gauge flux as shown in Fig.~\ref{Fig2}(d). The nodal line in $k_z = 0$ plane is not topologically unavoidable; however, its existence requires that all gauge flux in the $k_z = 0$ plane (except the loop itself ) must vanish.

The surface state of HgCr$_2$Se$_4$, i.e. Fermi arcs, are more stable than the accidental degeneracy in pyrochlore iridates \cite{Wan2011}, given that the band crossings of HgCr$_2$Se$_4$ are topologically unavoidable. Another feature is that the fermi arcs are interrupted by the $k_z = 0$ plane, where the nodal line exists. HgCr$_2$Se$_4$ is also a promising QAH material in its quantum well structure. When the well is thin enough, the band inversion in the bulk band structure will be removed entirely by the finite size effect. With increasing the thickness, finite size effect is getting weaker and the band inversion restores subsequently, leading to a quantized Hall coefficient $\sigma_{xy} = 2e^2/h$. In fact, the strong AHE in the bulk samples of HgCr$_2$Se$_4$ has already been observed \cite{Solin1997}. On the contrary, the AHE in pyrochlore iridates should be vanishing because of the AF configuration.

Inspired by the double-Weyl nodes in HgCr$_2$Se$_4$, Fang et al \cite{Fang2012} classified the two band crossing in $n$-fold rotational symmetric 3D system without TRS. By the $k\cdot p$ theory, they found that $C_{4,6}$ symmetry can support double-Weyl nodes on high-symmetry line, consistent with the above result in HgCr$_2$Se$_4$. Besides, the $C_6$ symmetry can also support triple-Weyl nodes, which carry the $\pm 3$ monopole charges and disperse cubically in the off-axis plane. If one change the magnetization direction from $(001)$ to $(111)$, the $C_4$ symmetry in HgCr$_2$Se$_4$ is broken whereas the rotation-reflection symmetry $S_6$ along $(111)$ direction arise. By calculating the $C_{3(111)} = S_6^2$ eigenvalues and the two-band $k\cdot p$ theory, Fang et al found that a double-Weyl node on $k_z$ axis with monopole charge $-2$ will evolve to a $+1$ Weyl node on $(111)$ axis and three $-1$ Weyl nodes off the axis related by $C_{3(111)}$ symmetry.

Although the experimental evidence has not been found yet, the prediction of a single pair of Weyl nodes in HgCr$_2$Se$_4$ inspired a series of works about the magnetic and transport properties of this material \cite{Guan2015,Lin2016HgCr2Se4,Lin2016CPL} and the quantum correction to the Hall conductance induced by electron-electron interaction \cite{Yang2019}. The transport studies on high quality HgCr$_2$Se$_4$ single crystals \cite{Guan2015} confirmed the spin-polarized current in its s-orbit conduction band, suggested its half-metal nature.

\subsection*{Magnetic Heusler Compounds}

In recent years, a series of papers predicted the Weyl nodes in Co$_2$-based magnetic full-Heusler compounds. Wang and collaborators \cite{Wang2016} studied Co$_2$XZ Heusler compounds (X=IVB or VB; Z=IVA or IIIA) and found that the favorable magnetization direction is along the [110] easy axis. In this configuration, there are at least two Weyl nodes close to the Fermi energy and largely separated in momentum space. {K\"ubler} and Felser \cite{Kubler2016} found that the large anomalous Hall effect in Co$_2$MnAl is possibly linked to its two pair of Weyl nodes, and suggested the same WSM phase for Co$_2$MnGa. {Soon after, Sakai et al \cite{Sakai2018} revealed the giant anomalous Nernst Effect in Co$_2$MnGa, and provided a guiding principle for increasing the intrinsic transverse thermoelectric conductivity. The Co$_2$MnGa compound is also predicted by Chang et al \cite{Chang2017b} to host the Hopf link protected by two perpendicular mirror plane, in which two nodal rings pass through the center of each other, and the Hopf link opens an extremely small gap ($<1mev$) under the SOC.}  Chang et al \cite{Chang2016} explored Co$_2$TiX (X=Si, Ge,or Sn) and found similar Weyl points in the [110] and [001] magnetization ground state.

\begin{figure*}[!h]
  \centering
  \includegraphics[width=1\linewidth]{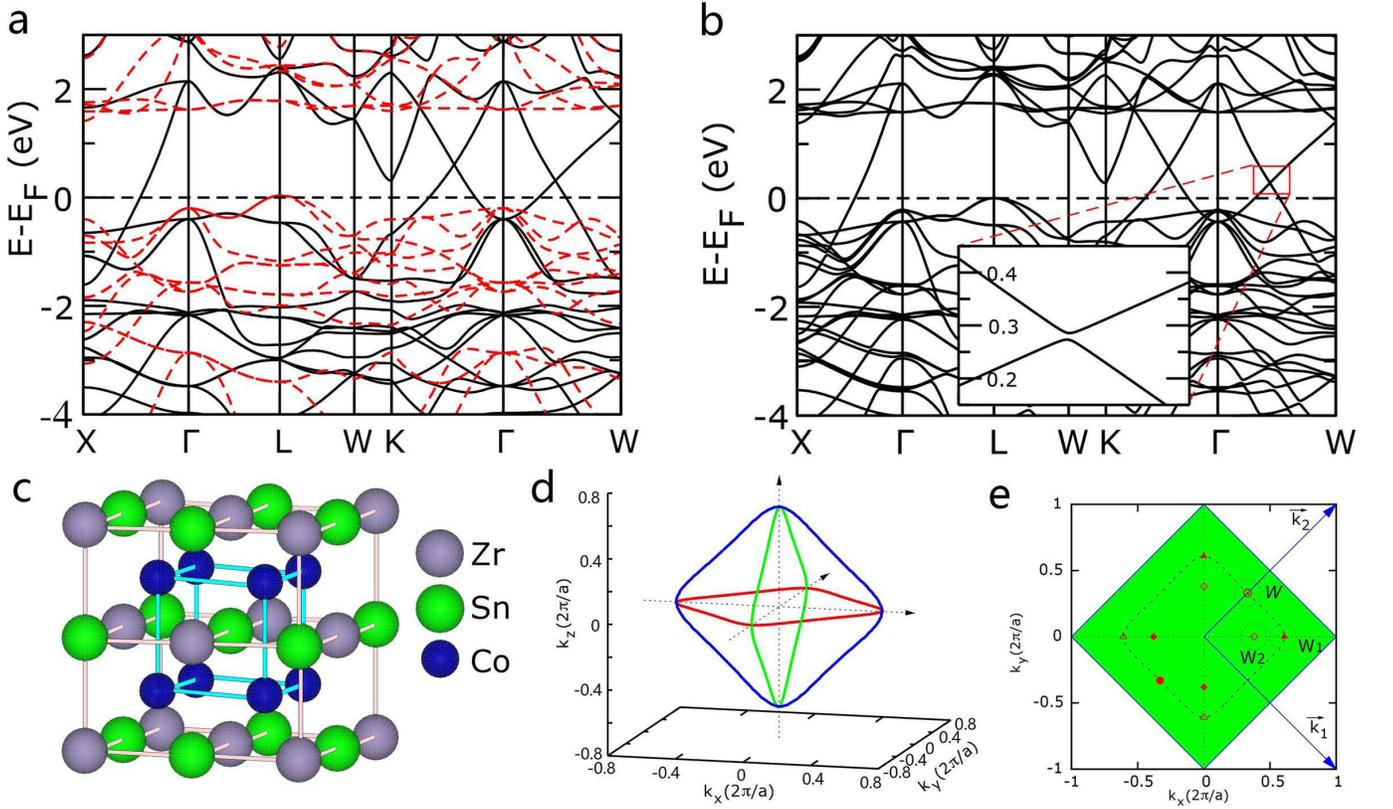}\\
  \caption{(a) The calculated band structure of Co$_2$ZrSn along high-symmetry lines without SOC. The majority and minority spin bands are denoted as solid-black and dashed-red lines, respectively. (b) The calculated band structure of Co$_2$ZrSn with SOC in the [110] magnetization configuration, the inset shows the small gap in the $\Gamma - W$ direction. (c) Rocksalt crystal structure of Co$_2$ZrSn. (d) Three nodal lines without SOC lie in $k_x-k_y$, $k_y-k_z$, $k_z-k_x$ planes and protected by $M_z$, $M_x$, $M_y$ respectively. (e) The chirality and position of Weyl nodes with SOC (top view), the remaining nodes be obtained by symmetry operation. W, W$_1$ and W$_2$ nodes are clearly independent, W and W$_1$ are in the $k_x-k_y$ plane, while the W$_2$ is out of the plane. With the Wilson-loop method, one can calculate the Chern numbers of a sphere enclosing a Weyl point to determine its chirality. The filled (unfilled) symbols indicate the chirality +1 (-1). The green square is the 001-surface BZ, with the surface lattice vectors $\vec{k_1}(2\pi/a,-2\pi/a)$ and $\vec{k_2}(2\pi/a,2\pi/a)$.\\
  (Figures are reprinted from ref. \cite{Wang2016}. Copyright \copyright2019 by the American Physical Society.)}\label{Fig3}
\end{figure*}
Full-Heusler are magnetic intermetallic compounds with face-centered cubic crystal structure X$_2$YZ (space group $Fm\bar{3}m$, No.225), with transition metal elements X, Y, and main-group element Z, with X the most electropositive \cite{Manna2018}. The proposed magnetic WSMs by Wang et al are Co$_2$XZ Heusler compounds (X=IVB or VB; Z=IVA or IIIA) with valence electrons number $N_v=26$, whose total spin magnetic moment $m=N_v - 24$, according to Slater-Pauling rule. Without loss of generality, it is convenient to focus on the candidate Co$_2$ZrSn, which has been synthesized experimentally \cite{Carbonari1996}, to discuss the topological semimetal phase. The GGA + U without SOC calculated spin-polarized band structure (Fig.~\ref{Fig3}(a)) reveals its half-metallic property, consistent with the experimental investigation of the spin resolved unoccupied DOS of the partner compound Co$_2$TiSn \cite{Klaer2009}. the partial DOS suggests the states near Fermi level are dominated by Co-d and Zr-d electrons. The SOC only has little influence on the band structure (Fig.~\ref{Fig3}(b)) and half-metallic property, because of the small SOC strength of both Co and Zr. The magnetization direction favors [110] and [100], the former is slightly lower than latter energetically. Both magnetism demonstrate topological phase with Weyl nodes and nodal lines, in the following, the magnetization is chosen along [110].

In the absence of SOC, the energy bands show three nodal lines in $xy, yz, zx$ plane, protected by their mirror symmetry $M_z,M_x,M_y$ respectively, as shown in Fig.~\ref{Fig3}(d). When considering SOC and in [110] spin polarization, some spatial crystal symmetry including $M_z,M_x,M_y$ are broken, leaving a magnetic space group generated by three elements: IS $I$, two fold rotation $C_{2[110]}$, and $C_{2z}T$ the combination of time reversal and $C_{2z}$. The nodal lines are gapped, except a pair of Weyl nodes survived along [110], protected by $C_{2[110]}$, i.e the crossing bands have different $C_{2[110]}$ eigenvalues $\pm i$ on the high-symmetry line. In addition, other two kinds of Weyl nodes can be found by carefully checking the nodal lines, as shown in Fig.~\ref{Fig3} (e). Four Weyl nodes (W$_1$) in $xy$ plane are related to each other by $I$ and $C_{2[110]}$, and eight general Weyl nodes (W$_2$) are related by all the three generators of the magnetic group. In fact, the product of the IS eigenvalues of the occupied bands at the inversion symmetric points is -1, hinting the presence of an odd number of pairs of Weyl nodes \cite{Hughes2011}. Those Weyl nodes position and topological charge and energy are presented in Table. \ref{Wang_table1}.
\begin{table}[]
  \centering
  \begin{tabular}{cccccc}
  \hline
      Compounds    & magnetization    & Weyl points    & coordinates    & Chern number    & $E-E_F$     \\
                   &                  &                & $(k_x \frac{2\pi}{a},k_y \frac{2\pi}{a},k_z \frac{2\pi}{a})$ &  & (eV) \\
    \hline
      Co$_2$ZrSn   & [110]            & W              & (0.334,0.334,0) & -1             & +0.6         \\
      (Co$_2$XZ)    &                  & W$_1$          & (0.58,-0.0005,0) & +1            & -0.6         \\
                   &                  & W$_2$         & (0.40,0.001,$\pm$0.28) & -1      & +0.3         \\
                   & [100]            & W          & (0.58,0,0)       & +2            & -0.6         \\
                   &                  & W$_1$         & (0.36,0.30,0)     & -1            & +0.55         \\
    \hline
      Co$_2$MnAl   & [110]            & W$_1$          & (0.5,0.81,0)    & +1             & +0.0309         \\
      (Co$_2$MnGa) &  [100]           & W$_1$          & (0.5,0.81,0)   & +1              & +0.0307         \\
    \hline
      Co$_2$TiGe   & [110]            & W          & $\sim$(0.408,0.408,0)    & +1             & +0.345         \\
      (Si,Sn)      &  [001]           & W$_1$          & (0,0,0.6)         & -2              & -0.285         \\
                   &                  & W$_2$         & (0,-0.29,0.46)       & +1            & +0.315         \\
                   &                  & W$_3$         & (0,-0.33,0.30)       & -1            & +0.315         \\
    \hline
  \end{tabular}
  \caption{Weyl nodes of Co$_2$ZrSn, Co$_2$MnAl and Co$_2$TiGe full-Heusler compounds in [110] and [100] magnetization direction respectively. Other Weyl nodes are related to the ones listed by the magnetic group ($I,C_{2z}T,C_{2[110]}$ for [110] magnetization, and $I,C_{4x},C_{2x}I,C_{2y}T,C_{2z}T$ for [100]). The data was taken from \cite{Wang2016,Kubler2016,Chang2016}}\label{Wang_table1}
\end{table}

W$_2$ Weyl nodes are removable by tuning SOC to move them to $k_z$ axis and annihilate; W$_1$ are locally stable in $k_z =0$ plane due to $C_{2z}T$ \cite{Soluyanov2015} but the energy of W$_1$ is very low; the Weyl nodes W, however, are topologically stable and can be tune to Fermi level by alloying. The 27-electrons Co-based Heusler family such as Co$_2$NbSn, which have also been synthesized experimentally \cite{Carbonari1996}, contains one more electron per a unit cell than that of Co$_2$ZrSn. Therefore, by alloying Co$_2$ZrSn with Nb in the Zr site, one can expect the Weyl nodes more close to Fermi level with the main band topology unchanged. The band structure calculation \cite{Wang2016} for Co$_2$Zr$_{1-x}$Nb$_x$Sn (with x = 0.275) shows that, in this concentration the Weyl nodes are bring to the Fermi level. For the other experimental synthesized 27-electron compound Co$_2$VSn \cite{Carbonari1996}, the alloy Co$_2$Ti$_{1-x}$V$_x$Sn (with x = 0.1) gives the same result.

When the magnetization parallel to [100] direction, the remained magnetic group is generated by: $I$, $C_{4x}$, $C_{2x}I$, $C_{2y}T$, $C_{2z}T$. Two $C_{4x}$ protected Weyl nodes with Chern number $\pm 2$ are found on $k_x$ axis. Due to the mirror symmetry $C_{2x}I$, the nodal line in $yz$ plane remains even with SOC. Also, $C_{2y}T$ ($C_{2z}T$) allows the existence of Weyl points in $xz$ plane ($xy$ plane), as shown in Table. \ref{Wang_table1}. The similar Weyl nodes have also been found in Co$_2$TiX (X=Si, Ge,or Sn) and Co$_2$MnAl(Ga) Heusler compounds, their coordinations, topological charges and energy to Fermi level are summarized in Table. \ref{Wang_table1}.

Comparing to other Weyl materials, magnetic Heusler compounds are ferromagnetic half-metal with Curie temperatures up to the room temperature \cite{Carbonari1996}, and their magnetism is ``soft'' and sensitive to external magnetic field. Chadov et al \cite{Chadov2017} studied the stability of the Weyl nodes in full-Heusler compounds, and found that number and coordinates of the Weyl nodes can be controlled by the magnetization direction. Moreover, the vast class of Heusler materials hints that one can tune those compounds across different compositions by alloying to get the desired properties. In summary, it is realistic to manipulate the spin and Weyl nodes in various of Heusler compounds, which provide a promising experimental platform to research spintronics and magnetic Weyl fermions.

\subsection*{Stacking Kagome Lattice}

One of the most exotic properties of magnetic WSM is the large intrinsic anomalous Hall effect, Which, in turn, provides a clue for magnetic WSM materials searching. {Very recently}, several reports proposed the existence of Weyl nodes in layered Kagome lattice \cite{Yang2017,Liu2018,Wang2018}. {Inspired by a series of {first principles} predictions \cite{Kubler2014,Chen2014,Zhang2017} and experimental discoveries \cite{Nakatsuji2015,Kiyohara2016,Nayake2016,Zhang2016} of AHE and spin Hall effect (SHE) in Mn$_3$X (X=Sn, Ge and Ir),} Yang et al \cite{Yang2017} confirmed the Weyl nodes in chiral anti-ferromagnetic Mn$_3$Sn and Mn$_3$Ge with Kagome layers Mn atoms by $ab\ initio$ calculation. On the other hand, 2D Kagome lattice with out-of-plane magnetization has become an excellent platform for AHE study \cite{Ohgushi2000,Xu2015prl115}. By stacking, it provides an effective way to realize magnetic WSMs \cite{Burkov2011,Zyuzin2012}. Following that guiding principle, two groups (Liu et al \cite{Liu2018} and Wang et al \cite{Wang2018}) individually claimed that out-of-plane magnetization Co$_3$Sn$_2$S$_2$ with Kagome layers Co atoms is a magnetic WSM candidate. These theoretical and experimental works suggests a new direction to search and synthesize magnetic WSMs among the materials with large AHE. Moreover, they will deepen our understanding on the microscopic mechanisms of the arising of AHE. In the following, we will introduce the theoretical result of Weyl nodes in Mn$_3$Sn (Mn$_3$Ge) and Co$_3$Sn$_2$S$_2$ in two sub-subsections, respectively.

\begin{figure}[!h]
  \centering
  \includegraphics[width=1\linewidth]{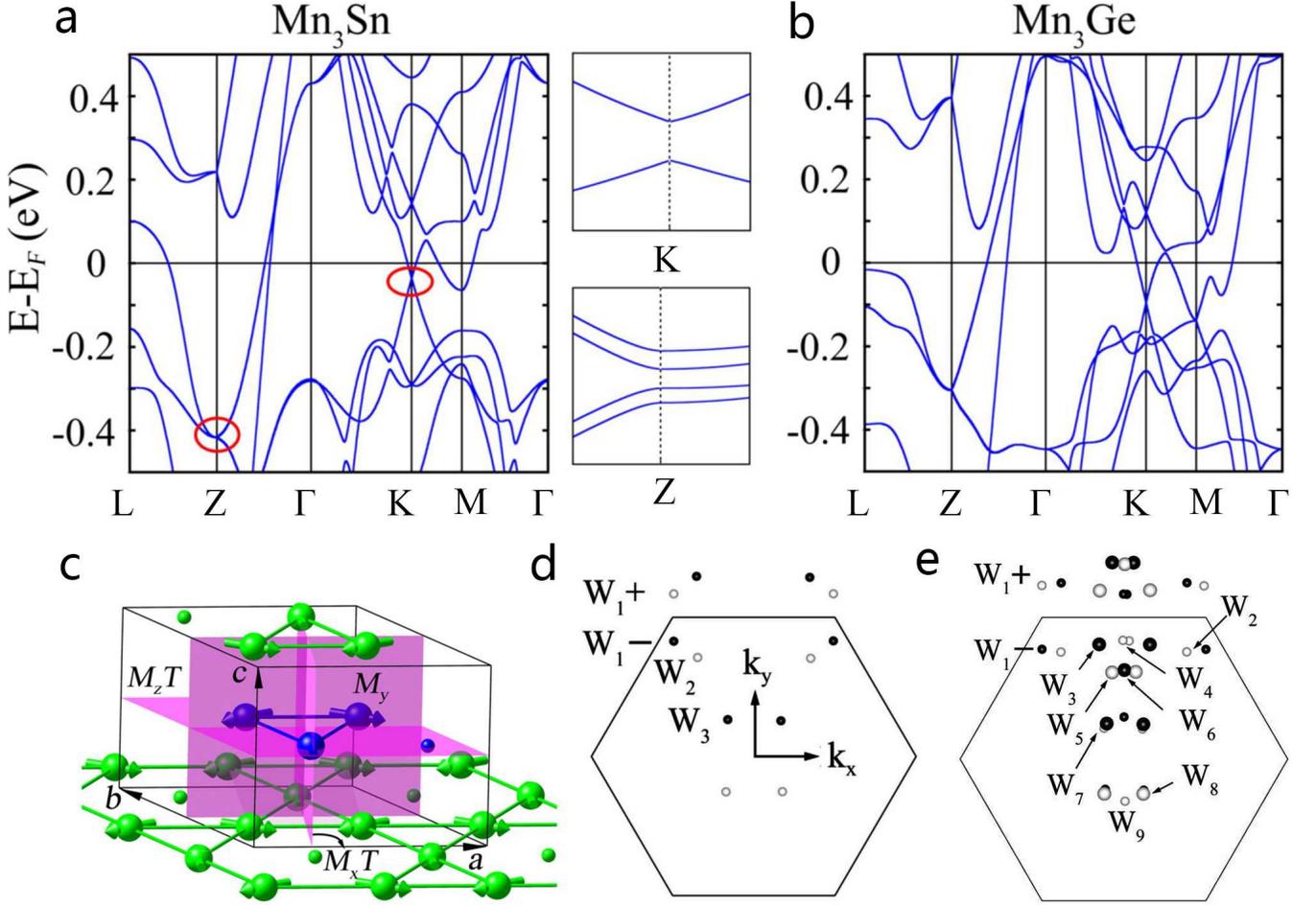}\\
  \caption{(a-b) The calculated band structures with SOC of (a)Mn$_3$Sn and (b)Mn$_3$Ge along high-symmetry lines. The energy gap near the $Z$ and $K$ (indicated by red circles) are shown in details. The Fermi energy is set to zero. (c) Crystal and magnetic structures of Mn$_3$X (X = Sn or Ge) and related magnetic mirror symmetry. The large (small) balls stand for Mn (X) atoms. The purple planes indicate three mirror planes of $\{M_y|\tau = c/2\}$,$M_xT$, and $M_zT$ symmetries respectively. (d-e) The chirality and position of Weyl nodes for (d)Mn$_3$Sn and (e)Mn$_3$Ge in momentum space. Black and white points represent Weyl nodes with $-1$ and $+1$ chirality, respectively. Larger points indicate two Weyl points ($\pm k_z$) projected into this plane. Figures are reproduced from ref. \cite{Yang2017}, CC BY 3.0 }\label{Fig4}
\end{figure}
\textit{Mn$_3$Sn (Mn$_3$Ge)}--In each layer of Mn$_3$Sn (Mn$_3$Ge) compound (space group $P6_3/mmc$, No.194), Mn atoms form a Kagome lattice with Ge(Sn) atoms located at the centers of each hexagons. In the ground magnetic states, Mn atom carries a magnetic moment of 3.2 $\mu B$ in Mn$_3$Sn (2.7 $\mu B$ in Mn$_3$Ge) and form a non-collinear AFM order. The magnetic moments lie inside the $xy$ plane with $2\pi/3$ angles between each two, as shown in Fig.~\ref{Fig4}(c). Such a non-collinear magnetic ground state is originated from the interplay of the easy-axis anisotropy and the SOC induced significant Dzyaloshinskii-Moriya (DM) interactions in the strongly frustrated kagome lattice \cite{Tomiyoshi1982,Tomiyoshi1983,Sticht1989,Brown1990,Sandratskii1996}.
This magnetic Kagome lattice has a nonsymmorphic symmetry $M_y\tau=\{M_y|0,0,1/2\}$ and two magnetic mirror {symmetries} $M_xT$ and $M_zT$.

Generally, the positions of Weyl nodes can be understood by symmetry analyzing. Time reversal operation will not change the chirality, while mirror reflection will reverse it. Hence, giving a Weyl node, other nodes related by $M_y\tau$, $M_xT$ and $M_zT$ will be settle down. However, the symmetries are slightly broken due to the tiny net moment in real materials ($\sim0.003\mu_B$ per unit cell). This weak symmetry broken is negligible for transport measurement, but will influence the band structure and induce a perturbation of the relationship of the Weyl nodes, for example, slightly shifting the positions of mirror partners, as shown in Table. \ref{Yang_table1}.

\begin{table}[]
  \centering
  \begin{tabular}{cccccc}
  \hline
      Compounds    & magnetization    & Weyl points    & coordinates    & Chern number    & $E-E_F$     \\
                   &                  &                & $(k_x \frac{2\pi}{a},k_y \frac{2\pi}{a},k_z \frac{2\pi}{a})$ &  & (eV) \\
    \hline
      Mn$_3$Sn   & non-collinear       & W$_1$          & (-0.325,0.405,0)      & -1             & +0.086         \\
                   & in-plane AFM     & W$_2$          & (-0.230,0.356,0.003)   & +1            & +0.158         \\
                   &                  & W$_3$         & (-0.107,0.133,0)         & -1           & +0.493         \\
    \hline
      Mn$_3$Ge   & non-collinear       & W$_1$          & (-0.333,0.388,-0.000)       & -1             & +0.057         \\
                &  in-plane AFM       & W$_2$          & (0.255,0.378,-0.000)        & +1              & +0.111         \\
                 &                    & W$_3$          & (-0.101,0.405,0.097)    & -1             & +0.048         \\
                    &                & W$_4$          & (-0.004,0.419,0.131)     & +1              & +0.008         \\
                   &                  & W$_5$         & (-0.048,0.306,0.164)       & +1            & +0.077         \\
                   &                  & W$_6$         & (0.002,0.314,0.171)       & -1            & +0.059         \\
                    &                & W$_7$          & (-0.081,0.109,0.000)         & +1              & +0.479         \\
                   &                  & W$_8$         & (0.069,-0.128,0.117)       & +1            & +0.330        \\
                   &                  & W$_9$         & (0.004,-0.149,-0.000)       & +1            & +0.470         \\
    \hline
  \end{tabular}
  \caption{Weyl nodes of Mn$_3$Sn and Mn$_3$Ge. Other Weyl nodes can be generated by the symmetries $M_y\tau$, $M_xT$ and $M_zT$. Noting that the symmetries are slightly broken by a tiny net magnetic moment, hence the coordinates will be slightly shifted from where they are expected to be. Table is reproduced from ref. \cite{Yang2017}, CC BY 3.0}\label{Yang_table1}
\end{table}
The bulk band structures with SOC of Mn$_3$Sn and Mn$_3$Ge exhibit similar dispersions, as shown in Fig.~\ref{Fig3}(a-b). At first glance, there are two seemingly band crossing points below the Fermi level at $Z$ and $K$. A tiny gap lifts the degeneracy and generates one pair of Weyl nodes near $Z$ and $K$ respectively. However, the Weyl node separations near $Z$ and $K$ are very small, and may generate negligible observable consequence in experiment. The physically interesting Weyl nodes are those general band crossings listed in the following.

In fact, Mn$_3$Sn and Mn$_3$Ge are metals with valence and conduction bands crossing many times near the Fermi level, leading to multiple pairs of Weyl nodes. Suppose the valence electron number is $N_v$ and count in the crossing between the $N_v^{th}$ and $(N_v+1)^{th}$ bands. In Mn$_3$Sn, there are 12 Weyl nodes classified into three groups (W$_1$, W$_2$, W$_3$, shown in Fig.~\ref{Fig4}(d) and Table. \ref{Yang_table1}, each one has three partners according to the symmetries). The Weyl nodes displayed in Mn$_3$Ge are more complicated, as shown in Fig.~\ref{Fig4}(e) and Table. \ref{Yang_table1}. There are nine groups of Weyl nodes with W$_{1,2,7,9}$ in the $k_z=0$ plane (W$_9$ also on the $k_y$ axis), W$_4$ in the $k_x=0$ plane, and $W_{3,5,6,8}$ in generic positions. Therefore, W$_{1,2,7,4}$ have other three partners, W$_9$ has other one partner, while $W_{3,5,6,8}$ have other seven partners according to the symmetries.

Right after the discovery of non-collinear magnetic WSM phase in Mn$_3$Sn and Mn$_3$Ge, Guo et al \cite{Guo2017} studied the large AHE, ANE as well as SHE and spin Nernst effect (SNE) in Mn$_3$X (X=Sn, Ge, Ga) through the $ab \ initio$ calculation of the Berry phase. The large AHE and the giant ANE in the non-collinear antiferromagnetic materials Mn$_3$Sn and Mn$_3$Ge can also be understood by the revised linear response tensor \cite{Seemann2015} and the cluster multipole extension method \cite{Suzuki2017,Suzuki2019}. The giant ANE has recently been experimentally confirmed by Ikhlas et al \cite{Ikhlas2017}, Li et al \cite{Li2017} and Kuroda et al \cite{Kuroda2017} in Mn$_3$Sn. Higo et al \cite{Higo2018} recently observed the large magneto-optical Kerr effect (MOKE) in Mn$_3$Sn. The interplay between the MOKE and the Fermi arcs caused by the Weyl nodes is an interesting question to be answer. The AHE induced by the Fermi arcs in the magnetic domain walls has been observed in Mn$_3$Sn(Ge) \cite{Liu2017,Li2019}. Recently, the proposed dynamics of the textures in the non-collinear antiferromagnets provide a theoretical mechanism for driving domain walls in Mn$_3$Sn(Ge, Ir) \cite{Yamane2019}, which is a platform to study the interplay between the magnetic Weyl nodes and the domain walls.

\begin{figure}[!h]
  \centering
  \includegraphics[width=0.7\linewidth]{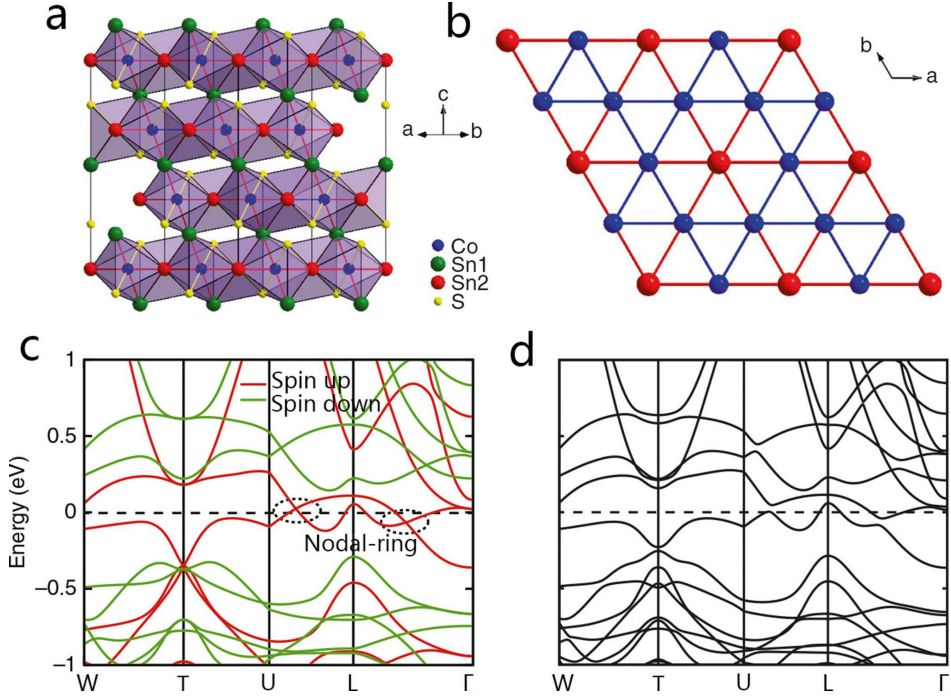}\\
  \caption{(a) The rhombohedral crystal structure of Co$_3$Sn$_2$S$_2$. Co and S atoms are represented by small blue and yellow balls, and the Sn atoms at Sn1 and Sn2 sites are represented by big green and red balls, respectively. (b) The Kagome layer formed by Co atoms. (c) The {first principles} calculated spin-resolved band structure without SOC. (d) The band structure with SOC. Figures are reproduced from ref. \cite{Wang2018}, CC BY 4.0}\label{Fig5}
\end{figure}
\textit{Co$_3$Sn$_2$S$_2$}--The structure of Co$_3$Sn$_2$S$_2$ compound is shown in Fig.~\ref{Fig5}(a-b), it is crystalized in a rhombohedral structure (space group $R\bar{3}m, No.166$) with a quasi-2D Co$_3$Sn layer sandwiched between sulfur atoms. The magnetic Co atoms form a perfect Kagome lattice in the $xy$ plane with ferromagnetic order along the easy $z$ axis (Curie temperature 177 K) and the magnetic moment is $0.29 \mu_B/$Co \cite{Richard2006,Vaqueiro2009,Schnelle2013}. The calculated band structure {by Wang et al \cite{Wang2018}} with and without SOC reveals the half-metallic feature with spin down gapped and spin up states crossing the Fermi level, consistent with the photoemission experimental measurements result \cite{Holder2009}.

When excluding SOC, there are linear band crossings along $\Gamma - L$ and $L - U$ line, as shown in Fig.~\ref{Fig5}(c). In fact, they are just single points of the nodal line in the mirror plane protected by the mirror symmetry $M_{y}$. According to the $C_{3z}$ and IS, there are six nodal lines in total in the BZ. Taking account the SOC, the mirror symmetry is broken. As a result, the nodal lines will be gapped as shown in Fig.~\ref{Fig5}(d), except three pairs of Weyl nodes off the high-symmetry line survived. Those Weyl nodes also related by $C_{3z}$ and IS, and contribute to the large intrinsic anomalous Hall effect in Co$_3$Sn$_2$S$_2$. Liu group  \cite{Liu2018} also reported the Weyl nodes induced negative magnetoresistance and large anomalous Hall angle, claimed that this ferromagnetic Kagome lattice is the first material hosting both a large anomalous Hall conductivity and a giant anomalous Hall angle that originate from the Berry curvature.

The Weyl nodes near the Fermi level means that Co$_3$Sn$_2$S$_2$ can host the large intrinsic transverse thermoelectric conductivity, and recently the giant ANE signal has been confirmed by Yang et al \cite{Yang2018a}.

\subsection*{GdSI}

Finding the systems exhibiting less pairs of Weyl nodes or other topological properties is a continuous mission. In 2017, Nie and corporators \cite{Nie2017} reported an IS broken honeycomb lattice model with promising topological phases and claimed that LnSI (Ln=Lu, Y, and Gd) satisfies this model. They predicted LuSI (YSI) as 3D strong TI, and GdSI can be an idea WSM with only two pairs of nodes.

\begin{figure}[!h]
  \centering
  \includegraphics[width=0.7\linewidth]{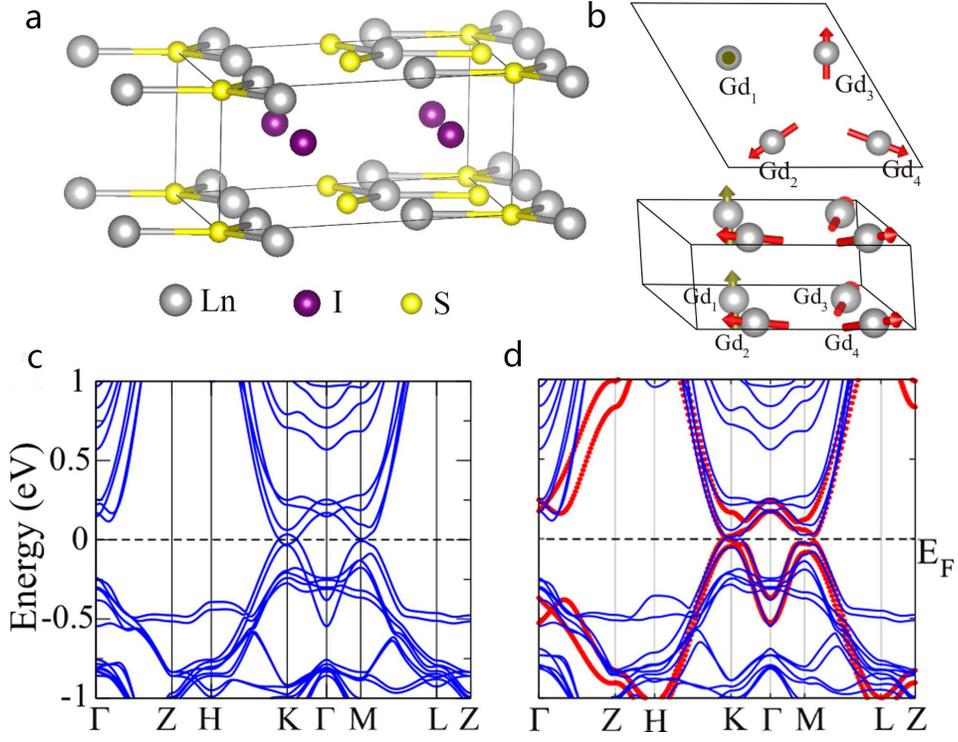}\\
  \caption{(a) Crystal structure of LnSI. Silver white, yellow, and purple balls represent Ln, S, and I atoms, respectively. (b) The top and side view of the non-collinear collinear magnetic configuration AFM4. (c-d) The band structures of GdSI calculated by GGA + U (b) and GGA + U + SOC (c), respectively. The fitted TB results are shown in (d) as red dots.}\label{Fig6}
\end{figure}
LnSI crystal has the space group $P\bar{6}$ \cite{Beck1986}, in which Ln atom and S atoms locate in the $xy$ plane to form a honeycomb lattice with I atoms intercalated between two LnS layers, see Fig.~\ref{Fig6}(a). The low energy bands near the Fermi level are dominated by the $p_z$ orbits of S atoms and the $d_{z^2}$ orbits of the Ln atoms. Although in each unit cell, there are four S and four Ln atoms, only one pair of $p_z$-type molecular orbital $|P_2\rangle$ with $j_z = \pm 1/2$ and one pair of $d_{z^2}$-type molecular orbital $|D_2\rangle$ with $j_z = \pm 1/2$ distribute to and invert at the Fermi level, owing to the chemical bonding and crystal field effects. For GdSI, the f orbits are partially occupied, hence GdSI is very likely to be stabilized in a magnetic phase. In fact, the GGA + $U$ + SOC method comparing different magnetic configurations shows that the most stable one is non-collinear collinear AFM4 as shown in Fig.~\ref{Fig6}(b), which breaks time reversal and the mirror symmetry $M_z$.

The calculated band structure reveals that GdSI is ideal WSM with two pairs of nodes (Fig.~\ref{Fig6}(c-d)). The band inversion  occurs near $\Gamma$ point and $K(K')$ point. Without SOC, due to the configuration II Rashba splitting, the Crossing bands belong to different eigenvalue of $M_z$, hence the crossings are stable and form nodal rings. However, SOC breaks $M_z$ symmetry and destroys nodal rings except two pairs of Weyl nodes on the high-symmetry $H-K(H'-K')$ line. They are protected by $C_{3z}$ symmetry due to the decrease of the effective angular momentum of $d_{z^2}$ orbits at $K$, which can be understood as following: without loss of generality, suppose Gd atom carrying $d_{z^2}$ is located at (1/3,2/3,0) in the honeycomb lattice and choose (0,0,0) as the rotation center. The rotation can be defined as $\hat{R}_3^z=e^{-i2\pi/3\hat{J}_z}$ with $\hat{J}_z = \hat{L}_z + \hat{S}_z$. Then one can get $\hat{R}_3^z |d_{z^2}^{\{1/3,2/3,0\}},j_z\rangle_K = e^{-i2\pi/3j_z} |d_{z^2}^{\{1/3,-1/3,0\}},j_z\rangle_K = e^{-i2\pi/3j_z}e^{i2\pi/3} |d_{z^2}^{\{1/3,2/3,0\}},j_z\rangle_K = e^{-i2\pi/3(j_z-1)} |d_{z^2}^{\{1/3,2/3,0\}},j_z\rangle_K$, where $K =(-1/3,2/3,0)$ is defined
with respect to the reciprocal lattice vectors. Therefore, the effective $j_z$ for the $d_{z^2}$ bands at K point will decrease by 1, becoming -1/2($|d_{z^2}\uparrow\rangle$) and -3/2($|d_{z^2}\downarrow\rangle$), respectively. However, the effective $j_z^K$ of the $p_z$ bands located at (0,0,0) site will not change.

The distribution of Weyl nodes in $k_z>0$ BZ is summarized in will have their counterparts at the same $k_x,k_y$ but opposite $k_z$, because the inverted bands are approximately symmetrical around $K$ ($K'$), despite the $M_z$ breaking in GdSI. The precise location of Weyl nodes given by the DFT calculation are $(-1/3,2/3,\pm 0.023)$ and $(1/3,-2/3,\pm 0.021)$, where the small difference between $K$ point and $K'$ point is induced by the TRS breaking.

\section*{\label{MDSM} Magnetic Dirac Semimetals}

A Dirac node is a four-fold degenerate point where two spin-degenerate bands cross. There are also some other Dirac nodes that we will not cover in this review, such as the double-refraction Dirac nodes, in which case the bands near the four-fold degenerate point will split to four non-degenerate bands. Generally, a DSM needs Kramers degeneracy at every $\vec{k}$-point to ensure double degeneracy everywhere in momentum space. In a nonmagnetic system, it needs the time-reversal $\mathcal{T}$ and inversion $I$ symmetries to be both preserved. In magnetic systems, $\mathcal{T}$ is broken, thus one may need a ``magnetic symmetry'', which is the product of a crystal symmetry with $\mathcal{T}$ to realize the Kramers degeneracy everywhere. The ``magnetic symmetry'' is often chosen as $I\mathcal{T}$ or $I\tau\mathcal{T}$, where $\tau$ is a slip operation. A space group containing a ``magnetic symmetry'' (anti-unitary generator) is called a magnetic space group (MSG).

The Dirac band crossing is not topologically stable. Generally, adding SOC can gap out the band crossing and change the Dirac node into the gapped dispersion relation of massive Dirac fermions. The material then becomes an insulator, which can be a topological insulator (TI), or a topological crystal insulator (TCI), etc. When the Dirac node is protected by crystal symmetry, e.g.~if the two two-fold-degenerate bands belong to different representations of some high-symmetry lines or points, the crossing is no longer avoided.
On the other hand, compared with the nonmagnetic Dirac semimetals, the magnetic Dirac quasiparticles can be controlled by the N$\acute{e}$el spin-orbit torques and induce the topological metal-insulator transition, in which, the N$\acute{e}$el vector orientation can switch on/off the symmetry that protect the Dirac band crossings \cite{Smejkal2017,Smejkal2017b}. Hence the TRS breaking Dirac semimetals are promising for the spin-orbitronics application \cite{Wadley2018,Smejkal2018,Schoop2018,Emmanouilidou2019}.
In the following, we review the prediction of magnetic DSMs CuMnAs and EuCd$_2$As$_2$, in which the ``magnetic symmetry'' causing Kramers degeneracy are $I\mathcal{T}$ and $I\tau\mathcal{T}$, and the Dirac nodes are protected by the screw axis $\tilde{C}_{2z} =\{C_{2z\,}|_{\,}\frac{1}{2}_{\,}0_{\,}\frac{1}{2}\}$ and three-fold rotation $C_{3z}$, respectively.

\subsection*{CuMnAs}

Magnetic DSM was firstly put forward by Peizhe Tang etc. \cite{Tang2016} and they proposed orthorhombic AFM CuMnAs as a candidate. Both TRS $T$ and IS $I$ are broken but their combination $IT$ is respected in this antiferromagnetic system, and screw rotational symmetry $\tilde{C}_{2z} = \{C_{2z}|\frac{1}{2}0\frac{1}{2}\}$ protected Dirac points are predicted to be robust on the high-symmetry X-U line ($k_x =\pi, k_y = 0$). A rough analyzation can be taken as following. First, the combination $IT$ symmetry gives the Kramers degeneracy everywhere. then to analyze the commuting relation between $IT$ and $\tilde{C}_{2z}$, one can denote $\tilde{C}_{2z}$ as $e^{i\pi \hat{j_z}}e^{ik_x/2+ik_z/2}$, then we have $(IT)\tilde{C}_{2z}(IT)^{-1} = \tilde{C}_{2z}e^{-ik_x - ik_z}$, which  becomes $(IT)\tilde{C}_{2z}(IT)^{-1} = -\tilde{C}_{2z}e^{ - ik_z}$ given $k_x = \pi$. Hence the Kramers pair states on high-symmetry line will have the same $\tilde{C}_{2z}$ eigenvalue, and if two pairs of bands crossing here have opposite $\tilde{C}_{2z}$ eigenvalue, i.e the different representation, the crossing is robust.

\begin{figure}[!h]
  \centering
  \includegraphics[width=0.9\linewidth]{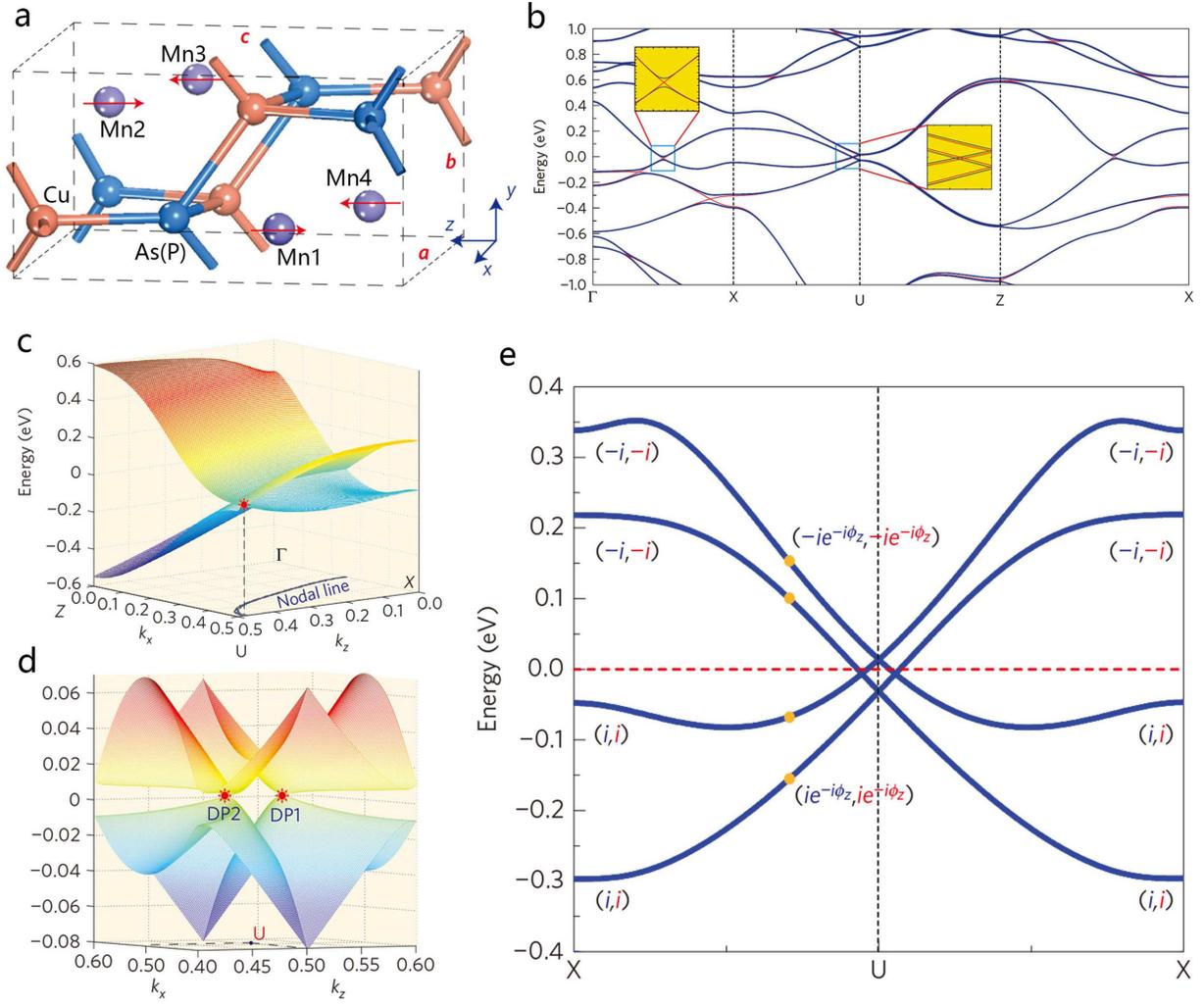}\\
  \caption{{(a) The crystal structure of the orthorhombic CuMnAs(P). The red arrows represent the orientations of magnetic moments on Mn atoms. (b) Calculated electronic structures of CuMnAs along the high-symmetric lines with (blue lines) and without (red lines) SOC. The magnetic moments of Mn atoms are chosen along the z direction when considering SOC. The insets show details of the band crossings near the Fermi level, which is set to be zero.} (c) The calculated band structure of orthorhombic CuMnAs around the Fermi level with $R_y$ preserved. The black line illustrated the Dirac nodal line. (d) The calculated band structure of orthorhombic CuMnAs with $R_y$ broken by shear strain. The red stars illustrated the Dirac nodes around the Fermi level. (e) The calculated band structure of orthorhombic CuMnAs with SOC along the high-symmetry line $X-U-X$. The orientation of magnetic moments is chosen to along $z$ direction. $\pm i$ are the eigenvalues of the screw rotation symmetry $\tilde{C}_{2z}$ at the $X$ point. Along the high-symmetry line, the eigenvalue of $\tilde{C}_{2z}$ is $\pm ie^{-i\phi_z}$. The blue and red colours stand for different spin states, respectively.\\
  (Figures are reprinted from ref. \cite{Tang2016}, Springer Nature. \copyright2016 by Springer Nature Customer Service Centre GmbH.)}\label{Fig7}
\end{figure}
CuMnAs and CuMnP have already been confirmed experimentally as room-temperature antiferromagnets \cite{Maca2012,Wadley587}, where non-zero magnetic moments of $3d$ electrons on Mn atoms order anti-ferromagnetically, see Fig.~\ref{Fig7}(a). Their crystal structure has the non-symmorphic space group $D_{2h}(Pnma)$ with four formula units in the primitive unit cell. This space group has eight symmetry operations and can be generated by the IS $I$, and two non-symmorphic symmetries: the gliding mirror reflection of the $y$ plane $R_y = \{M_y|0\frac{1}{2}0\}$, and the two-fold screw rotation along the $z$ axis $\tilde{C}_{2z} = \{C_{2z}|\frac{1}{2}0\frac{1}{2}\}$. Considering the magnetic configuration will break some symmetries. In the most energy-favoured AFM configuration in the orthorhombic phase, the magnetic moments on the inversion-related Mn atoms are aligned along opposite directions, which breaks both $T$ and $I$ but preserves $IT$. If SOC is absent, the internal spin space is decoupled from real space, hence the spatial symmetries $R_y$ and $\tilde{C}_{2z}$ are kept. While when considering SOC, the residual symmetries will depend on the orientation of magnetic moments. For example, only $\tilde{C}_{2z}$ can survive if magnetic moments are along the z direction, and protect Weyl nodes on high-symmetry line $X-U$, as shown in Fig.~\ref{Fig7}(b).

The {first principles} calculated band structure are shown in Fig.~\ref{Fig7}(c) for a case where SOC is turned off in the antiferromagnetic system.One can find band crossings along high-symmetry lines, which are consistent with the previous report \cite{Maca2012}. Beyond these crossings, one can also find an entire elliptic Dirac nodal line (DNL) on the $k_y = 0$ plane around the Fermi level and centred at the X point. Examining the band dispersions under various perturbations shows that no gap opening along the nodal line as long as $R_y$ is present. Nevertheless, the nodal structure is not protected by $R_y$ because $R_y$ and $IT$ commute on the $k_y = 0$ plane which gives the fact that Kramers pair here have opposite $R_y$ eigenvalue. By checking the orbital composition of the bands, one will confirm that the existence of such a DNL without SOC is associated with the behaviours of the underlying atomic orbits under $R_y$. For one of the crossing bands, it is composed by $d_{xy}$ and $d_{yz}$ orbits that are odd under the mirror reflection, while the the other band is composed by $d_{xz}$,  $d_{z^2}$ and $d_{x^2-y^2}$ orbits that are even under the mirror reflection. The hopping terms between them must vanish, therefore the gapless DNL is strongly depends on the detailed electronic structures around the Fermi level. Corresponding to the DNL in the bulk, dispersive drumhead-like surface state will appear inside the projection of the DNL on the $(010)$ surface. Such a nontrivial surface state can be measured as a clear signature of the DNL semimetal \cite{Kim2015} \cite{Yu2015}.

If one still exclude SOC but break $R_y$ and keep $\tilde{C}_{2z}$ symmetry by, for example, applying the shear strain and shift the Mn atoms, the DNL will open a band gap except at four discrete points. One pair of them are located on the high-symmetry X-U line (Fig.~\ref{Fig7}(d)), and the other pair is located in the interior of the Brillouin zone. The first pair of four-fold degenerate points are verified to be Dirac points and are guaranteed by the screw rotation symmetry $\tilde{C}_{2z}$. Unlike $R_y$, $\tilde{C}_{2z}$ and $IT$ are anti-commutative along the X-U line, thus the doubly degenerate states at each $k$ point along this line have the same $\tilde{C}_{2z}$ eigenvalue. therefore, as long as the pair of doubly degenerate bands carry different $\tilde{C}_{2z}$ eigenvalue, their crossing must be stable. Based on the { $ab\ initio$ } results, the calculated $\tilde{C}_{2z}$ eigenvalues of the bands near the Fermi level exactly match the symmetry argument. The other pair of Dirac points in the interior of BZ are enforced by Nielsen-Ninomiya theory \cite{Nielsen1981a}. The argument is that a Dirac point without SOC is made up by two opposite Weyl points and each Weyl point have definite spin. For either spin components, the chirality of the Weyl points on the X-U line are found to be the same. As a result, other two Weyl points carrying opposite chirality must exist in the BZ to vanish the total chirality.

When SOC is turned on, some crystalline symmetries can be broken by the magnetism, therefore the stability of the crossing points sensitively depends on the orientation of the Mn atoms' local magnetic moments. If they are aligned along the z axis, only $\tilde{C}_{2z}$ symmetry from the space group survives. As shown in Fig.~\ref{Fig7}(e), in this case, the symmetry argument above for the robust crossing points on the X-U line still holds, hence the four-fold degenerate points here are intact under the protection of $\tilde{C}_{2z}$, while the other pair of crossing points are fully gapped. If the magnetic moments are along other directions, $\tilde{C}_{2z}$ is broken generally, and the Dirac fermions will obtain mass terms proportional to the strength of SOC. For orthorhombic CuMnAs and CuMnP considered here, the typical energy dependence on the magnetic moments orientation is relatively weak; therefore, to realize stable massless Dirac fermions here, several feasible methods, such as via proximity coupling \cite{Katmis2016}, can be taken to pin the moments along the $z$ axis even at finite temperatures.

Similar to non-magnetic Dirac and WSMs, the nontrivial surface arc state and the orbital texture of Dirac cones could be the direct evidence for the magnetic Dirac fermions. And since the net magnetization in CuMnAs and CuMnP are zero, the arc state and orbital texture can be measured by ARPES \cite{Xu2015347,Xu2015349}. Large spin Hall effects could appear in the Dirac fermions system, in which these relativistic particles could contribute to electric control of local magnetization in $IT$ invariant anti-ferromagnets. Although the magnetic configuration in the calculation is assumed to be frozen, in fact, AFM fluctuations are inevitably present in CuMnAs and CuMnP. In the massive  Dirac fermions, the fluctuations act as the dynamical axion field and cause the exotic modulation of the electromagnetic field \cite{Li2010}.
All discussion in this subsection is based on the local moments totally along $z$ axis. The moments along other direction, and the AFM fluctuation both may break the crystal symmetries that protect the band crossing, and lead to a massive Dirac fermion behaviour in this system.
The interplay between Dirac fermions, the AFM fluctuations and the symmetry breaking is still under research. Its exact description remains an open question.

\subsection*{EuCd$_2$As$_2$}

Hua et al \cite{Hua2018} exhaustively analyzed the DSMs in the magnetic space groups (MSGs), and proposed a candidate, the inter-layer AFM EuCd$_2$As$_2$, as a DSM in centrosymmetric type-IV MSGs, where the group $\mathcal{M}$ are defined as $\mathcal{G} + T\tau \mathcal{G}$.

\begin{figure}[!h]
  \centering
  \includegraphics[clip, width=0.7\linewidth]{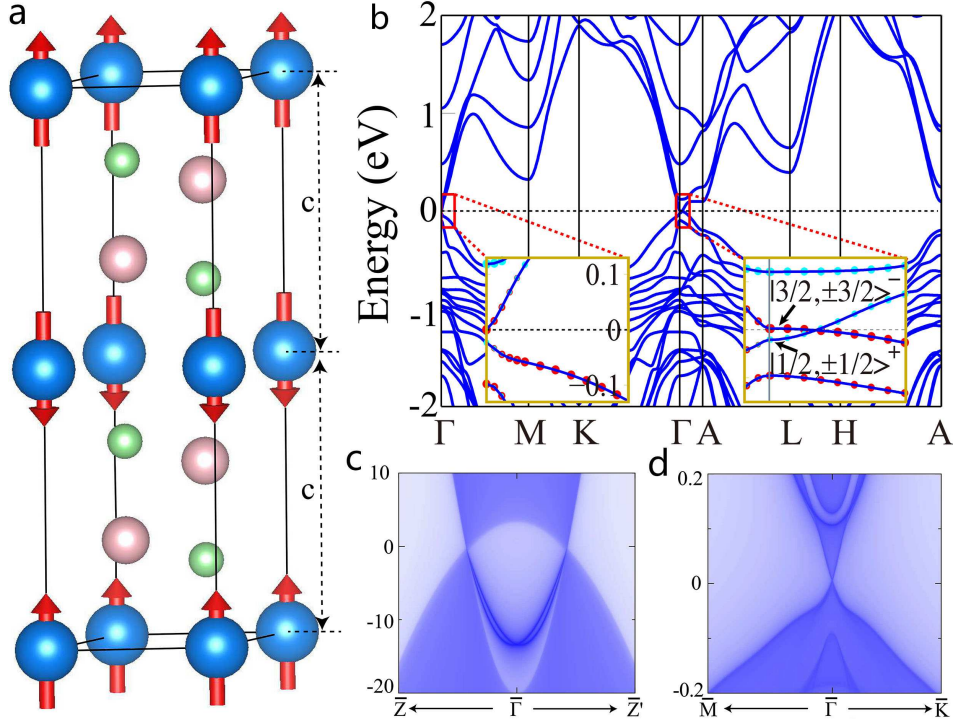}\\
  \caption{(a) Crystal structure of the inter-layer AFM EuCd$_2$As$_2$. The blue, pink and light green balls indicate Eu, Cd and As atoms, respectively. The red arrows represent the directions of the magnetic momentum. (b) The band structures of the inter-layer AFM EuCd$_2$As$_2$ calculated by GGA+U+SOC method. The insets are the zoom-in of the band structures around the $\Gamma$ point to clearly show the band inversion and Dirac node. The red and light blue dots demonstrate the projections of the As $p$ and Cd $s$ orbits, respectively. (c) and (d) are the calculated surface states on the (100) and (001) faces, respectively.\\
  (Figures are reprinted from ref. \cite{Hua2018}. Copyright \copyright2019 by the American Physical Society.)}\label{Fig8}
\end{figure}
As shown in Fig.~\ref{Fig8}(a), EuCd$_2$As$_2$ crystallizes into the CaAl$_2$Si$_2$-type structure (space group $P\bar{3}m1$, No.164) \cite{Artmann1996,Inga2009} with Cd$_2$As$_2$ layers separated by the trigonal Eu layers. Eu$^{2+}$ has a half-filled 4f shell, and the inter-layer AFM magnetic configuration is the most stable one. Fig.~\ref{Fig8}(b) shows the projected band structures of the inter-layer AFM EuCd$_2$As$_2$, where the low energy bands near the Fermi level are mainly contributed from the $p$ orbits of As atoms and the $s$ orbits of the Cd atoms. Around the $\Gamma$ point, the doubly degenerate $s-s$ bonding states of Cd atoms (even parity) invert with the $p-p$ anti-bonding states of As atoms (odd parity), causing a Dirac band crossing along $\Gamma - A$ line protected by $C_{3z}$ symmetry.

A detailed symmetry analysis reveals that a nonsymmorphic TRS $T' = T \oplus c$, connecting the up-spin momentum layer at $z = 0$ and the down-spin momentum layer at $z = c$, exists in this inter-layer AFM system. The MSGs of the inter-layer AFM EuCd$_2$As$_2$ can be expressed as $D_{3d}^4 \oplus T'D_{3d}^4$, generated by $T'$, IS $I$, rotation symmetry $C_{3z}$ and twofold screw $\tilde{C}_{2x} = C_{2x} \oplus c $. Combining $T' = T \oplus c$ with $I$, the anti-unitarity of $PT'$ would prohibit the hopping between the nonsymmorphic time reversal pair of states, such as $|3/2,\pm 3/2\rangle$ or $|1/2,\pm 1/2\rangle$, hence every energy state is doubly degenerate in such an inter-layer AFM system.

Along $\Gamma - A$ line, the little group can be described as $C_{3v} \oplus PT'C_{3v}$. When SOC is included, the topology and band inversion of the system are dominated by the four states: $|3/2,\pm 3/2\rangle^-$ from the $p-p$ anti-bonding states of As and $|1/2,\pm 1/2\rangle^+$ from the $s-s$ bonding states of Cd. The $4 \times 4$ effective $k\cdot p$ Hamiltonian around $\Gamma$ points, under symmetry restrictions, can be written as (in the order of $|1/2, 1/2\rangle^+$, $|3/2, 3/2\rangle^-$, $|1/2, -1/2\rangle^+$, $|3/2,-3/2\rangle^-$)
\begin{equation}\label{Hua_equ1}
  H= \epsilon_0(k) + \left(
                       \begin{array}{cccc}
                         M(k) & Ak_+ & 0 & Bk_+ \\
                         Ak_- & -M(k) & Bk_+ & 0 \\
                         0 & Bk_- & M(k) & -Ak_- \\
                         Bk_- & 0 & -Ak_+ & -M(k) \\
                       \end{array}
                     \right)
\end{equation}
where $\epsilon_0(k) = C_0 + C_1k_z^2 + C_2(k^2_x + k^2_y)$, $k_\pm = k_x \pm ik_y$ and $M(k) = M_0 - M_1k^2_z - M_2(k^2_x + k^2_y)$ with $M_0,M_1,M_2 < 0$ to guarantee the band inversion. This effective model is very similar to the Hamiltonian in Na$_3$Bi, except that the off diagonal terms here is the leading order $Bk_\pm$ rather than high order $Bk_zk_\pm^2$ as in Na$_3$Bi. There are two double degenerate eigenvalues $E_\pm = \epsilon_0 \pm \Delta$ with $\Delta = \sqrt{(A^2 + B^2)(k^2_x + k^2_y) +M^2(k)}$, which tells two linear Dirac nodes at $k_c = (0,0,\pm\sqrt{M_0/M_1})$ along the $\Gamma - A$ line. The Dirac nodes are confirmed by the calculated surface states and Fermi arcs shown in Fig.~\ref{Fig8}(c) and (d) based on the semi-infinite Green's functions constructed by the maximally localized Wannier functions \cite{Sancho1984,Sancho1985}. The (001) surface states shown in Fig.~\ref{Fig8}(d) exhibit a clear band touching at the $\Gamma$ point and Fermi level, where two Dirac nodes are projected to the same point. Moreover, a pair of Fermi arc states unambiguously connect the Dirac nodes on the (100) face as plotted in Figs.\ref{Fig8}(c). Even though the Fermi arcs appear to be closed, their Fermi velocities are discontinuous at the Dirac nodes.

Such a AFM DSM has its own uniqueness. Such a uniqueness is reflected by its derivatives, which makes it different from Na$_3$Bi and CuMnAs. When $C_{3z}$ symmetry is broken, $j_z$ is no longer a good quantum number, as a result, the hopping terms between $|j_z = \pm1/2\rangle$ and $|j_z = \pm3/2\rangle$ can be introduced, and the system will evolve into a strong TI phase due to the inverted band structure. However, due to the nonsymmorphic TRS $T' = T \oplus c$, the boundary states will be gapped on (001) surface, where $T'$ symmetry is broken. Therefore, a nontrivial AFM $Z_2$ invariant protected by $T'$ can be defined, and the half-quantum Hall effect can be realized on the intrinsically gapped (001) face of such an AFM TI \cite{Mong2010}. On the other hand, when $I$ symmetry is broken, instead of splitting into two pairs of ordinary Weyl points, the AFM DSM will split into two pairs of triple points protected by the small $C_{3v}$ group. This is due to that the magnetic point group $C_{3v}$ on $\Gamma - A$ has one 2D irreducible representation $E_{1/2}$ ($|\pm 1/2\rangle$) and two one-dimensional irreducible representations $E_{3/2}$ ($\frac{1}{\sqrt{2}}|3/2\rangle \pm \frac{i}{\sqrt{2}}|-3/2\rangle$). Hence, the degeneracy between $|\pm 3/2\rangle$ states originally protected by $PT'$ is broken, while the degeneracy between $|\pm 1/2\rangle$ remains, naturally leading to two pairs of triple points along the $\Gamma - A$ line.

\section*{\label{MNLSM} Magnetic Nodal Line Semimetals}

Nodal line semimetals (NLSMs) can be viewed as having a line of Weyl nodes or Dirac nodes with no dispersion along the nodal line and linear dispersion in perpendicular directions. Similar to the 1D Fermi-arc surface states in WSMs, NLSMs have the nearly dispersionless 2D ``drumhead'' surface states embedded inside the band gap between the conduction and valence bands in the 2D projection of the nodal ring, and the drumhead states have infinite DOS. Like Dirac nodes, NLSMs are not topologically stable and need crystalline symmetries to protect the band crossing. When the protecting symmetry breaks, the nodal line can be either fully gapped or gapped into several nodes. So analyzing the evolution of the nodal line is helpful to predict new topological insulators or semimetals. For more details, one can read the recent review articles by Fang et al  \cite{Fang2016} and Yang et al  \cite{Yang2018}. So far the research of NLSMs is mainly on time-reversal-preserved systems with or without SOC. Several modelling works have studied the magnetic NLSMs via symmetry analysis \cite{Burkov2011PRB,Wang2017PRB,Yu2017PRB}, and generally the {first principles} predicted magnetic NLSMs emerge as the intermediate phase of a magnetic WSM and a magnetic DSM, as the nodal lines evolve into discrete nodes under SOC. For example, the band structure of GdSI without SOC shows nodal rings protected by $M_z$, which evolve into two pairs of Weyl nodes due to the breaking of $M_z$ symmetry by SOC. There are also SOC-immune but impure (coexist with nodal points) magnetic NLSMs predicted. In the ferromagnetic material HgCr$_2$Se$_4$  \cite{Xu2011HgCr2Se4}, the mirror symmetry $M_z$ protects the magnetic Weyl nodal line in the $k_z = 0$ plane, which coexists with the Weyl nodes on the $k_z$-axis. In Co$_2$-based Heusler compounds with $(100)$ magnetization configuration  \cite{Wang2016,Kubler2016,Chang2016}, the nodal lines in the $k_y = 0$ and $k_z = 0$ planes are gapped into Weyl nodes while the nodal line in the $k_x = 0$ plane remains intact due to the preserved $M_x$ symmetry under SOC. So far, finding pure magnetic NLSMs with stable nodal lines near the Fermi level is still on the way, especially in the presence of SOC. Recently, Kim et al  \cite{Kim2018} proposed Van der Waals material Fe$_3$GeTe$_2$ and Nie et al \cite{Nie2019} proposed the 3D layered LaX (X=Cl, Br) as candidates for ferromagnetic NLSMs that are robust against SOC, which will be reviewed in the rest of this section.

\subsection*{Fe$_3$GeTe$_2$}

\begin{figure*}[!h]
  \centering
  \includegraphics[width=1\linewidth]{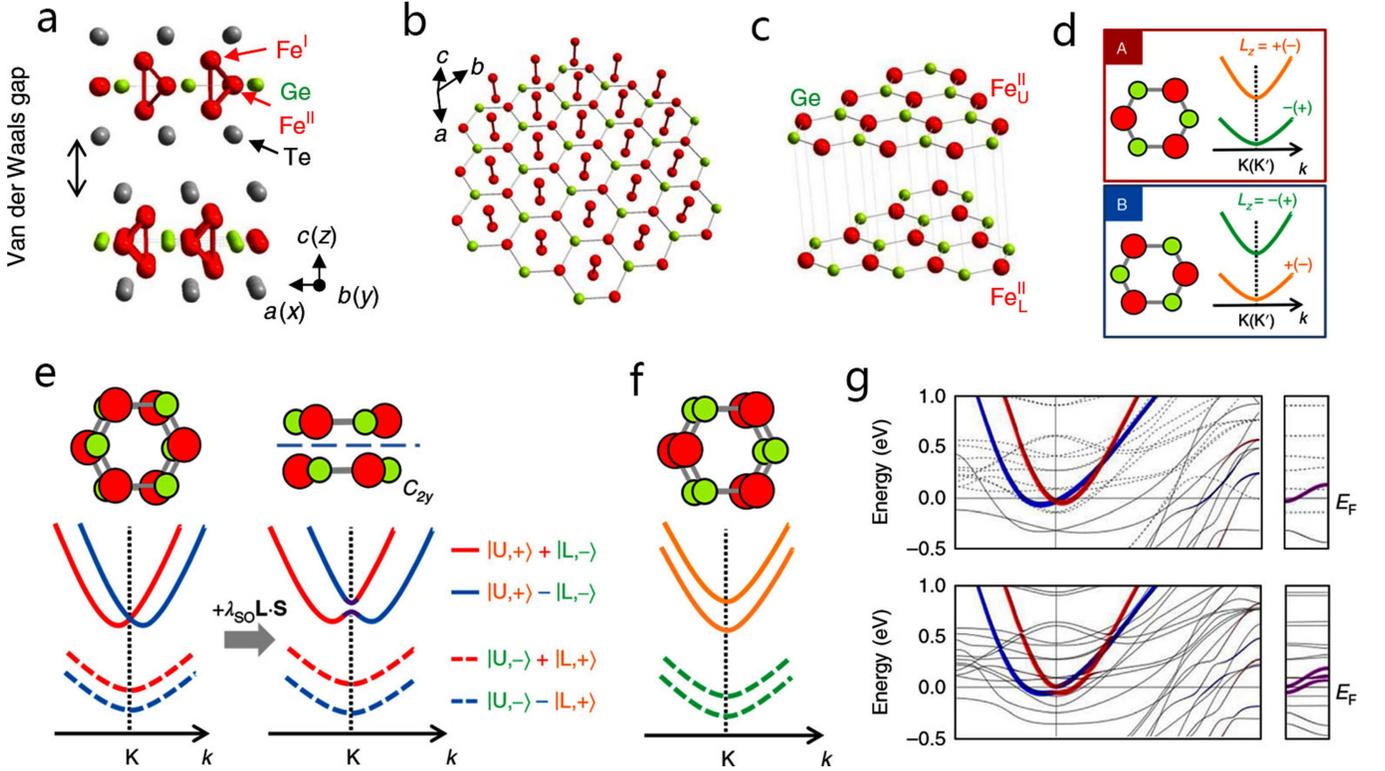}\\
  \caption{Crystal structure of Fe$_3$GeTe$_2$ and the band crossing. (a) Structure of a Fe$_3$GeTe$_2$ bilayer. (b) Structure of a Fe$_3$Ge monolayer. (c) Simplified structures of a Fe$_3$GeTe$_2$ bilayer with only the Fe$^{II}$ and Ge atoms. (d) Different orbital splitting at the $K$ ($K'$) points for A and B types atomic configurations, respectively. (e) The band structure at the $K$ point for an AB stacked bilayer with inter-layer hybridization. The orbital-driven band degeneracy at $K$ ($K'$) point is protected by $C_{2y}$, which can be lifted by the $\lambda_{SO}\mathbf{L}\cdot\mathbf{S}$. The orbital degrees of freedom are denoted by $\pm$, and layers are denoted by U or L. (f) The band structure for an AA stacked bilayer. (g) Calculated band structures of Fe$_3$GeTe$_2$ without (upper panel) and with (lower panel) SOC. The solid (dashed) lines stand for majority (minority) spin, and the colours indicate two orthogonal states with opposite orbital and layer degrees of freedom.\\
  (Figures are reprinted from ref. \cite{Kim2018}, Springer Nature. \copyright2018 by Springer Nature Customer Service Centre GmbH.)}\label{Fig9}
\end{figure*}

In 2018, Kim et al \cite{Kim2018} predicted and ARPES detected the Van der Waals material Fe$_3$GeTe$_2$ as a candidate ferromagnetic NLSM, which is stable in the case that orbital and spin angular momenta are perpendicular. The layered Fe$_3$GeTe$_2$ is an itinerant electron ferromagnetic material with the Curie temperature high to $T_c = 220K$ \cite{Chen2013,Andrew2016}. In the hexagonal crystal structure as shown in Fig.~\ref{Fig9}(a), the Fe$_3$Ge slabs are coupled via vdW interaction, and sandwiched by Te layers. The Fe$^I$-Fe$^I$ pairs across the centre of the hexagonal plaquettes in the covalently bonded Fe$^{II}$-Ge honeycomb lattice (Fig.~\ref{Fig9}(b)). The AB stacking configuration (Fig.~\ref{Fig9}(c)) of Fe$^{II}$-Ge layers is essential for the fully spin polarized nodal line degeneracy because of its particular crystalline symmetry.

Before describing the AB stacking Fe$^{II}$-Ge layer structure, one can first consider the Fe$^{II}$-Ge bilayer as shown in Fig.~\ref{Fig9}(c), which has the space group $P\bar{3}1m$ (No.164) generated by $C_{3z}$, $C_{2y}$ and the IS $I$. The high-symmetry $K$ point is invariant under $C_{3z}$, $C_{2y}$ and $IT$, which allows 2D irreducible representation and therefore, double degeneracy in the absence of SOC. Given that the bands are fully spin polarized, the degeneracy comes from the orbital degree of freedom, as illustrated in Fig.~\ref{Fig9}(d) and (e). On the contrary, in the AA stacking hypothetical bilayer (Fig.~\ref{Fig9}(f)), $C_{2y}$ and $I$ are broken and the new symmetry $C_{2x}$ is not a relevant symmetry at the $K$ point. As a result, the $C_{3z}$ symmetry at $K$ point only have 1D irreducible representation characterized by its eigenvalues, and hence the bands are non-degenerate here.

The 3D vdW layered Fe$_3$GeTe$_2$ structure belongs to the space group $P6_3/mmc$ (No.194), which is generated by $\tilde{C}_{6z} = \{C_{6z}|00\frac{1}{2}\}$, $C_{2y}$ and $I$. The high-symmetry point $K$ ($H$) has little group generated by $\tilde{C}_{6z}I$, $C_{2y}$ and $IT$, which allows 2D irreducible representations. Without SOC, the calculated band structure confirms the crossing at $K$ point and the typical Mexican-hat due to Rashba effect, as shown in Fig.~\ref{Fig9}(g). The crossing bands come from the mixed $3d$ orbits with Fe$^I$-Fe$^I$ $L_z = \pm 1$ states and Fe$^{II}$ $L_z = \pm 2$ states. More specifically, the eigenstates crossing at $K$ can be denoted by $\psi_{1,k} = |L_z=+1\rangle^I_U + |L_z=+2\rangle^{II}_U$ and $\psi_{2,k} = |L_z=-1\rangle^I_L + |L_z=-2\rangle^{II}_L$. In this bases, the symmetry operators $\tilde{C}_{6z}I$ and $C_{2y}$ can be represented as $\tilde{C}_{6z}I = \cos{\frac{2\pi}{3}}+i\sin{\frac{2\pi}{3}}\tau_z$ and $C_{2y} = \tau_x$ with the Pauli matrices $\tau_{x,y,z}$ denoting $\psi_{1,k}$ and $\psi_{2,k}$. The effective $k\cdot p$ Hamiltonian can be written as
\begin{equation}\label{Kim_eq1}
  H_0 = \epsilon_0 + \frac{1}{2m_{xy}}(k_x^2+k_y^2) + \frac{1}{2m_z}k_z^2 + \alpha [k_y\tau_x + k_x\tau_y]
\end{equation}
with the parameters $\epsilon_0 = -0.03$~eV, $m_{xy} = 0.077$~eV$^{-1}{\AA}^{-2}$, $m_z = 0.036eV^{-1}{\AA}^{-2}$ and $\alpha = 0.71eV{\AA}$ estimated from the band calculation. straightforwardly, there is band degeneracy along $K-H$ line, and also $K'-H'$ line according to $\tilde{C}_{6z}$ symmetry. The nodal lines are protected by $C_{3z}$ and $\tilde{C}_{6z}M_y$ ($IT$), which gives the 2D irreducible representations along $K-H$ line.

Now consider the influence of SOC with the form $H_{SO} = \lambda_{SO}\mathbf{L}\cdot\mathbf{S}$, which in the ferromagnetic spin configuration, can be treated as $H_{SO} \approx \lambda_{SO}\mathbf{L}\cdot\langle\mathbf{S}\rangle$. The crossing bands are composed of $|L_z = \pm 1\rangle^I$ and $|L_z = \pm 2\rangle^{II}$ states, hence in the bases of $\psi_{1,k}$ and $\psi_{2,k}$, $\langle L_x \rangle = \langle L_y \rangle = 0$ and $\langle L_z \rangle = \frac{4}{3}\tau_z$. As a result, the nodal lines are stable even with SOC in the situation of $\mathbf{S}\parallel x(y)$. On the other hand, SOC will open a gap of $\sim 60$meV along the nodal line if $\mathbf{S}\parallel z$. Generally, the SOC gap depends on the angle between $\mathbf{S}$ and $z$ direction. Therefore, in this ferromagnetic NLSM Fe$_3$GeTe$_2$, the stability of the band crossing and its SOC gap can be tuned by the fully spin polarization direction.

Unfortunately, the real Fe$_3$GeTe$_2$ magnetic configuration favors the spin polarization along $z$ direction, thus the nodal lines will open a gap in the presence of SOC and produce a Berry flux along the line to induce the large AHE. Hence, to find a proper material with magnetic nodal lines which is robust even against SOC is still an urgent task in the topological semimetal field.

\subsection*{LaCl (LaBr)}

In 2019, Nie et al  \cite{Nie2019} predicted the spinful nodal lines in 3D layered materials LaX (X=Cl, Br), which are constructed by stacking 2D Weyl materials. Generally, Weyl nodes can exist in 2D materials protected by crystal symmetry, for example, one pair of Weyl nodes protected by mirror symmetry $M_y$. When the 2D WSM is stacked into 3D layered system, three classes of topological semimetals can be obtained according to the symmetry and the inter-layer coupling strength. Class 1 is two nodal lines extending through the BZ when the inter-layer coupling is weak. Class 2 is nodal loops or nodal chains with strong inter-layer coupling. Class 3 is 3D WSM if the symmetry on the stacking line is broken. Following the guideline, Nie et al found the idea 2D WSMs LaX, and due to the weak inter-layer coupling, the 3D layered LaX are ferromagnetic NLSMs with a pair of nodal lines extending through the BZ and protected by mirror symmetry.

\begin{figure*}[!h]
  \centering
  \includegraphics[width=1\linewidth]{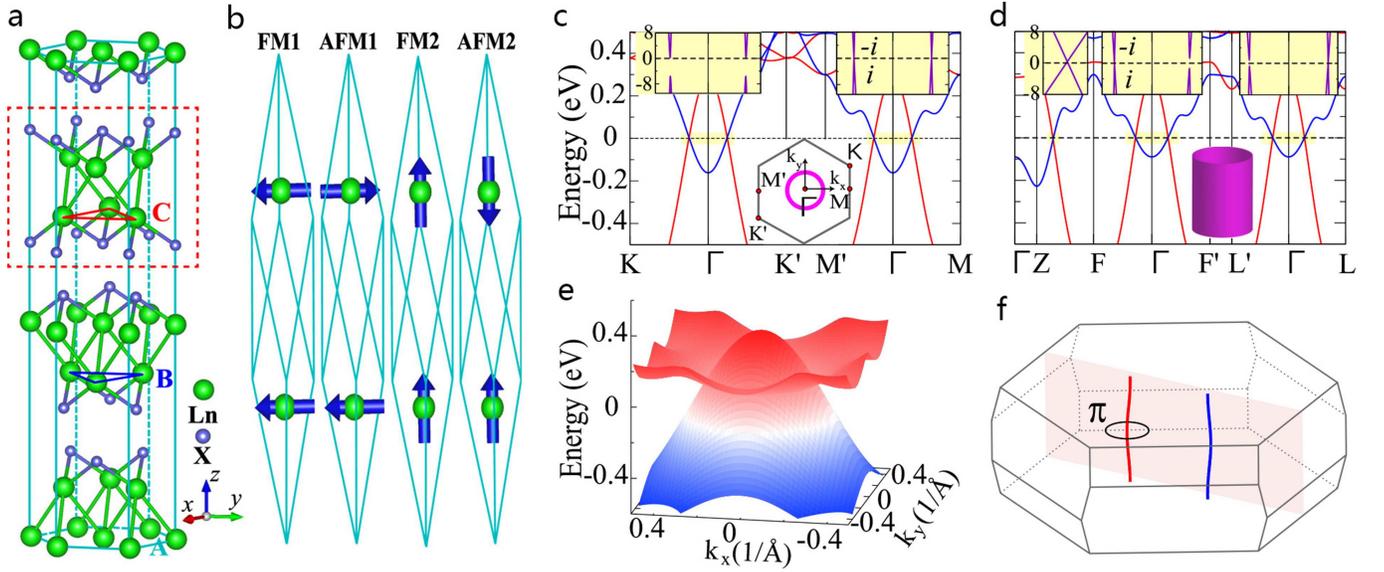}\\
  \caption{(a) Crystal structure of 3D layered material LaX (X=Cl, Br). The quadruple layer is marked by the red dashed line. (b) Schematic illustration of four different collinear magnetic configurations in a primitive cell of LaX. Only magnetic La atoms are drawn, and X atoms are omitted for simplicity. (c-f) Calculated band structures of LaX. (c,d) The band structures of single-layer LaCl and 3D LaCl by LDA +U. The red (blue) lines denote the spin-up (down) states. The upper insets are the zoom-in band structure with SOC around the band crossing near Fermi level, and their energy unit is $meV$ instead of $eV$. $\pm i$ here are the eigenvalues of mirror symmetry $M_y$. The lower insets schematically show the nodal line (c) and cylinder (d), respectively. (e) The band structures of 3D LaCl in the $k_x-k_y$ plane without SOC. (f) Schematic illustration of two nodal lines for 3D LaCl in the $k_x-k_z$ plane.\\
  (Figures are reprinted from ref. \cite{Nie2019}. Copyright \copyright2019 by the American Physical Society.)}\label{Fig10}
\end{figure*}
The LaX crystal has the hexagonal layered structure (space group $R\bar{3}m$, No.166) as shown in Fig.~\ref{Fig10}(a). It is built by stacking the tightly bound quadruple layer, which has X-La-La-X sub-layers made up by two hexagonal rare-earth-metal La layers sandwiched between two hexagonal halogen (X) layers. The stacking pattern is ABC-type trilayer along the $z$ axis with weak van der Waals interaction. The inter-layer distance is around $10{\AA}$ and the inter-layer coupling is much weaker than most layered compounds, including graphene. As a result, it is easy to obtain the 2D single-layer LaX through exfoliation methods. The calculated total energies of different magnetic configurations (see Fig.~\ref{Fig10}(b), ferromagnetic FM1, FM2 and antiferromagnetic AFM1, AFM2) for 2D and 3D LaX suggest FM1 where the easy magnetization axis lies in the $xy$ plane. Due to the almost negligible magneto-crystalline anisotropy (the energy is of the order of 0.0001meV), the spin prefers to align along the $y$ direction.

For the 2D single-layer LaCl (LaBr will have the almost same result), the calculated band structures reveal a deep inversion at $\Gamma$ point, as shown in Fig.~\ref{Fig10}(c). Without SOC, the band inversion forms a nodal line around $\Gamma$ point. With the consideration of SOC, the nodal line opens a gap except two Weyl nodes on the $M-\Gamma-M'$ line, which are protected by mirror symmetry $M_y$. When the idea 2D WSMs are stacked to build a 3D LaCl, due to the extremely weak inter-layer coupling, one may obtain a 3D nodal line semimetal (class 1). In fact, the calculated band structures of 3D LaCl confirm this speculation as shown in Fig.~\ref{Fig10}(d-f). Without SOC, the nodal line from single-layer LaCl forms a cylinder centered around the $\Gamma$ point, as schematically shown in the lower inset of Fig.~\ref{Fig10}(d). When considering SOC, the cylinder is gapped out everywhere except two nodal lines in the $k_x-k_z$ plane crossing through the BZ (class 1). The crossing two bands have opposite eigenvalue of $M_y$ as shown in the upper inset of Fig.~\ref{Fig10}(d), hence the nodal lines are protected by $M_y$ and robust to SOC.

Different from the ordinary nodal lines, the nodal lines in LaCl (LaBr) always appear in pairs because of the IS $I$. One pair of nodal lines can meet and annihilate in the momentum space without breaking the mirror symmetry $M_y$. On the other hand, the 3D LaX (X=Cl, Br) ferromagnetic NLSM with one pair of spinful nodal lines extending through the BZ is so far the only predicted magnetic NLSM candidate exactly robust against SOC. This discovery is meaningful to open a new path to realize and research the long pursued nodal-line fermions, and its interplay with magnetization.

\section*{\label{Prospects} Discussion and Outlook}

Topological semimetals extend the topological classification of materials from insulators to metallic systems, and have become one of the most attractive fields of study in condensed matter physics in recent years. Nonmagnetic TSMs have been well studied theoretically and the mapping from the topological classification of TRS-preserved materials to the band representation of the space group at high-symmetry points in BZ has been well established \cite{Bradlyn2017,Po2017,SongNC2018,SongPhysRevX.8.031069,Zhang2018}. The database of topological materials including nonmagnetic TSMs has been published by Weng \textit{et al.} (see http:// \linebreak materiae.iphy.ac.cn/). On the other hand, magnetic TSMs are a relatively new area that is still far from well developed. Although the topological classification of magnetic insulators has been performed based on the co-representation theory and K-homology of magnetic point groups \cite{Zhang2015,Okuma2018}, the topological phases of magnetic TSMs based on magnetic space groups are awaiting further studies. Magnetic DSMs and NLSMs that are robust against SOC are presently quite rare. In the search of new magnetic TSMs, {first principles} calculations have encountered big challenges, as they often overestimate the energy gain of magnetization and end up predicting wrong magnetic configurations and direction. On the experimental side, conventional ARPES measurements are usually ineffective to study the band structures of magnetic TSMs because of the magnetic domain wall problem. Most experimentally reported magnetic WSMs are based on indirect evidences such as large AHE and anomalous Nernst effect. Nevertheless, magnetic TSMs have their unique advantages. Because their symmetries and electronic structures depend sensitively on their magnetic structures and direction, it becomes convenient and realistic to manipulate their topological properties and phase transition by applying an external magnetic field, which can be highly useful in the design of spintronic devices. Hence, more research works are necessary to find out robust and high-quality magnetic TSMs with the degenerate points or lines close to Fermi level. More reliable experimental methods are needed to measure and confirm the topological properties of magnetic TSMs, such as the Weyl nodes, drumhead surface state etc.

``New fermions'' such as those in TPSMs can also be realized in magnetic materials. In the antiferromagnetic EuCd$_2$As$_2$, when inversion symmetry is broken, a Dirac node will evolve to a pair of three-fold degenerate nodes, each protected by the $C_{3v}$-symmetry. Cheung et al \cite{Cheung2018} have checked all the magnetic symmorphic point groups to search for triple points protected on high-symmetry line and found that Dirac and triple points can coexist in particular systems. The nonsymmorphic antiferromagnetic CeSbTe is predicted to host various topological states including Dirac and Weyl as well as triple and eight-fold degenerate points \cite{Schoop2018}. A full classification and understanding of the topological properties in the MSGs is far from completion, and is still an open question waiting for answers, in which completely new topological states, beyond all the known topological states at present, may be discovered.

Finally, electron-electron correlations are usually very important in magnetic materials. The interplay of topological order with electron-electron correlations remains a widely open question \cite{Leon2010,Nagaosa2016}.

\emph{Acknowledgements} The authors thank the support by the Ministry of Science and Technology of China (2018YFA0307000), and the National Natural Science Foundation of China (11874022); G. X. is supported by the National Thousand-Young-Talents Program.

\emph{Contributions} J.Z and G.X contributed to the collection of references and outline of the review paper. Z. H participated in the mathematical formalisms and interpretations of the basic theory. All authors contributed to writing the manuscript.

\emph{Competing interests} The authors declare no competing interests.

\bibliography{MTSM}

\end{document}